\documentclass[aps,prd,nofootinbib,preprintnumbers,twocolumn,showpacs, floatfix]{revtex4-1}
\usepackage{amsmath}
\usepackage{amssymb}
\usepackage{epsfig}
\usepackage{graphicx}
\usepackage{multirow}
\usepackage{color}
\usepackage[normal]{subfigure}
\usepackage{rotating}
\usepackage{hyperref}

\newcommand{\capdef}{}
\newcommand{\mycaption}[2][\capdef]{\renewcommand{\capdef}{#2}
\caption[#1]{{\footnotesize #2}}}

\newcommand{\ewt}{\end{widetext}}
\newcommand{\be}{\begin{equation}}
\newcommand{\ee}{\end{equation}}
\newcommand{\bdm}{\begin{displaymath}}
\newcommand{\edm}{\end{displaymath}}
\newcommand{\bea}{\begin{eqnarray}}
\newcommand{\eea}{\end{eqnarray}}
\newcommand{\nn}{\nonumber}

\def\eq#1{{Eq.~(\ref{#1})}}
\def\eqs#1#2{{Eqs.~(\ref{#1})--(\ref{#2})}}
\def\fig#1{{Fig.~\ref{#1}}}

\def\Table#1{{TABLE~\ref{#1}}}

\def\sect#1{{Sect.~\ref{#1}}}

\def\citelist#1#2{{[\citenum{#1}--\citenum{#2}]}}
\def\vev#1{\left\langle #1 \right\rangle}

\def\Tr{\mbox{Tr}\,}

\begin{document}
\preprint{TTP13-007, SFB/CPP-13-14}
%----------------------------------------------------------------------------------
\title{Light color octet scalars in the minimal SO(10) grand unification}
%----------------------------------------------------------------------------------
\author{Stefano Bertolini}\email{bertolin@sissa.it}
\affiliation{INFN, Sezione di Trieste, SISSA,
via Bonomea 265, 34136 Trieste, Italy}
\author{Luca Di Luzio}\email{diluzio@kit.edu}
\affiliation{Institut f\"{u}r Theoretische Teilchenphysik,
Karlsruhe Institute of Technology (KIT), D-76128 Karlsruhe, Germany}
\author{Michal Malinsk\'{y}}\email{malinsky@ipnp.troja.mff.cuni.cz}
\affiliation{Institute of Particle and Nuclear Physics,
Faculty of Mathematics and Physics,
Charles University in Prague, V Hole\v{s}ovi\v{c}k\'ach 2,
180 00 Praha 8, Czech Republic}
%----------------------------------------------------------------------------------
\begin{abstract}
We analyze the relation between the present (and foreseen) bounds on matter stability and the presence of TeV-scale color octet scalar states in 
nonsupersymmetric $SO(10)$ grand unification with one adjoint Higgs representation triggering the symmetry breaking. 
This scenario, discarded long ago due to tree-level tachyonic instabilities appearing in all phenomenologically viable breaking patterns, 
has been recently revived at the quantum level.~By including the relevant two-loop corrections we 
find a tight correlation between the octet mass and the unification scale which either requires a light color octet scalar within the reach of the LHC 
or, alternatively, a proton lifetime accessible to the forthcoming megaton-scale facilities. \\
\end{abstract}
\maketitle
%----------------------------------------------------------------------------------

%%%%%%%%%%%%%%%%%%%%%%%%%%%%%%%%%%%%%%%%%%%
\section{Introduction}
%%%%%%%%%%%%%%%%%%%%%%%%%%%%%%%%%%%%%%%%%%%
The class of $SO(10)$ models where, the first step of the  spontaneous symmetry breaking is driven by the vacuum expectation value (VEV) of a Higgs adjoint, was for a long time  considered to be phenomenologically impracticable due to instabilities of the classical vacuum configurations supporting potentially viable symmetry breaking chains~\cite{Yasue:1980fy,Anastaze:1983zk,Babu:1984mz,Li:1973mq}. 
However, it was shown recently~\cite{Bertolini:2009es} that such instabilities  can be removed at the quantum level and, thus, the nonsupersymmetric
$SO(10)$ unification with the minimal Higgs content has been revived as a potentially realistic framework.

In a recent work~\cite{Bertolini:2012im} we analyzed a simple nonsupersymmetric $SO(10)$ gauge model with $45_H \oplus 126_H$ in the Higgs sector focusing on the details of the scalar spectrum as a key ingredient of a detailed understanding of the gauge unification constraints.~Surprisingly, the theory turned out to be capable of supporting viable unifying patterns with the $B-L$ breaking scale stretching as high as $10^{14}$ GeV, right within the ballpark assumed for a natural implementation of the seesaw for neutrinos (see~\cite{Bertolini:2009qj} for a systematic discussion).

In particular, we pointed out that a seesaw scale in the $10^{13}-10^{14}$ GeV range is obtained along the breaking chains featuring intermediate $SU(3)_c \otimes SU(2)_L \otimes SU(2)_R \otimes U(1)_{B-L}$ or 
$SU(4)_C \otimes SU(2)_L \otimes U(1)_R$ gauge symmetries when either an intermediate-scale color
sextet (transforming as a weak triplet) or a light color octet (weak doublet) appear in the scalar 
spectrum; the latter case is especially interesting because there the unification constraints allow for a colored scalar octet in the vicinity of the electroweak scale which can be very interesting from the collider physics point of view. On top of that, the mass of the octet turns out to be anticorrelated with the masses of the grand unified theory (GUT) scale vector bosons governing the $d=6$ proton decay and, hence, a lower bound on the matter lifetime translates to an upper bound on the octet mass.

From the bottom-up perspective, the existence of light color octet (or sextet) scalars in the sub-TeV domain was recently 
advocated in~\cite{Dorsner:2012pp,Reece:2012gi} as a possible explanation for the $H\to\gamma\gamma$ excess observed in the LHC 
data \cite{ATLAS:2012gk,CMS:2012gu,ATLAS:2012uma}. 
To this end, it has been pointed out~\cite{Manohar:2006ga} that only
the standard Higgs doublets or color octets with the Higgs-like 
weak quantum numbers can naturally
couple to quarks without introducing large flavor changing
neutral currents (FCNC). The implementation of a custodial symmetry in these settings can then help further with taming the associated  
radiative corrections so that a standard-model-like (SM-like) setup is maintained.
Light colored scalars have been widely discussed in connection to various new physics scenarios and in the recent years with great emphasis on their implications for the LHC physics \citelist{Frampton:1987dn}{Calvet:2012rk}.

In this paper we restrict the systematic discussion of ref.~\cite{Bertolini:2012im} to the possibility of a light color octet scalar. We refine the analysis of gauge unification to the two-loop level and we enter the details of the calculation of the width of the dominant proton decay mode. 
We entirely focus on the minimal $SO(10)$ light color octet (weak doublet) scenario since, unlike the intermediate-scale sextet case (with the typical sextet mass at about $10^{11}$ GeV), it is far more interesting from the collider perspective. 

We find that the two-loop corrections strengthen the
correlation between the scalar octet mass and the unification scale in such a way
that strict bounds on the scalar mass can be derived from the proton lifetime limits. In particular, for the present day proton stability
constraints we find a conservative 2000~TeV upper bound on the color octet mass which is further pushed down to about $20$~TeV in case of a null result in the next generation proton decay searches. In any case, a large fraction of
the parameter space of the minimal $SO(10)$ GUT allows for a scalar octet mass in the TeV (and even \nolinebreak{sub-TeV}) regime.

%%%%%%%%%%%%%%%%%%%%%%%%%%%%%%%%%%%%%%%%%%%
\section{Light color octet scalar in the minimal SO(10)}
%%%%%%%%%%%%%%%%%%%%%%%%%%%%%%%%%%%%%%%%%%%
\label{45126Higgsmodel}
\subsection{The 45-126 Higgs model}
We consider an $SO(10)$ Higgs sector including one $45_H$ adjoint representation together
with one $126_H$\footnote{Minimally, one would prefer to consider $16_H$ in place
of $126_H$. On the other hand $\vev{16_H}$ breaks the $B-L$ symmetry only by one unit and, thus, the seesaw requires a pair of  $\vev{16_H}$ insertions. This can be implemented at the renormalizable level by, e.g.,~a variant of the Witten's radiative mechanism \cite{Witten:1979nr,Bajc:2004hr,Bajc:2005aq} or,  giving up renormalizability, by a $d=5$ operator. In either case the ``effective'' $\Delta(B-L) = 2$ seesaw scale is further suppressed with respect to the ``genuine'' $B-L$ breaking VEV and the  light neutrino masses are typically overshot by many orders of magnitude.}. The consideration of an additional 
$10_H$ (or two) is not relevant for the purpose
of this paper and we shall comment on it only later in Sect.~\ref{MSH}. 

The most general renormalizable scalar potential that can be written with just $45_H$ and $126_H$ at hand reads
\be
\label{scalpotgen}
V = V_{45} + V_{126} + V_{\rm mix} \, ,
\ee
where
\bea
\label{V45}
V_{45} &=& - \frac{\mu^2}{2} (\phi \phi)_0 + \frac{a_0}{4} (\phi \phi)_0 (\phi \phi)_0 + \frac{a_2}{4} (\phi \phi)_2 (\phi \phi)_2 \, , \\ \nn \\
\label{V126}
V_{126} &=&  - \frac{\nu^2}{5!} (\Sigma \Sigma^*)_0\\
& +& \frac{\lambda_0}{(5!)^2} (\Sigma \Sigma^*)_0 (\Sigma \Sigma^*)_0 
+ \frac{\lambda_2}{(4!)^2} (\Sigma \Sigma^*)_2 (\Sigma \Sigma^*)_2\nn\\
& + & \frac{\lambda_4}{(3!)^2(2!)^2} (\Sigma \Sigma^*)_4 (\Sigma \Sigma^*)_4
+ \frac{\lambda'_{4}}{(3!)^2} (\Sigma \Sigma^*)_{4'} (\Sigma \Sigma^*)_{4'} \nn \\ \nn \\
&+& \frac{\eta_2}{(4!)^2} (\Sigma \Sigma)_2 (\Sigma \Sigma)_2
+ \frac{\eta_2^*}{(4!)^2} (\Sigma^* \Sigma^*)_2 (\Sigma^* \Sigma^*)_2 \, , \nn \\ \nn \\
\label{V45126}
V_{\rm mix} &=& \frac{i \tau}{4!} (\phi)_2 (\Sigma \Sigma^*)_2 
+ \frac{\alpha}{2 \cdot 5!} (\phi \phi)_0 (\Sigma \Sigma^*)_0 \\
%+ \beta_2 (\phi \phi)_2 (\Sigma \Sigma^*)_2
&+& \frac{\beta_4}{4 \cdot 3!} (\phi \phi)_4 (\Sigma \Sigma^*)_4
+ \frac{\beta'_{4}}{3!} (\phi \phi)_{4'} (\Sigma \Sigma^*)_{4'} \nn \\ \nn \\ 
&+& \frac{\gamma_2}{4!} (\phi \phi)_2 (\Sigma \Sigma)_2
+ \frac{\gamma_2^*}{4!} (\phi \phi)_2 (\Sigma^* \Sigma^*)_2 \,\nn .
\eea
Here the $\phi$ and $\Sigma$ stand for the components of $45_H$ and $126_H$, respectively. The detailed breakdown of all the contractions (with the subscripts denoting the number of open indices in the relevant brackets) is given  in Appendix A of ref.~\cite{Bertolini:2012im}. 
All couplings are real but $\eta_2$~and~$\gamma_2$.

There are, in general, three SM singlets in the reducible $45_H\oplus126_H$ representation of $SO(10)$. 
Using $BL \equiv (B-L)$
and labeling the field components with respect to the
$3_{c}\, 2_{L}\, 2_{R}\, 1_{BL}$ (i.e.,  $SU(3)_c \otimes SU(2)_L \otimes SU(2)_R \otimes U(1)_{BL}$) algebra, the SM singlets
reside in the $(1,1,1,0)$ and $(1,1,3,0)$ submultiplets of $45_{H}$
and in the $(1,1,3,+2)$ component of $126_{H}$.
In what follows we shall denote
\be
\label{vevs}
\vev{(1,1,1,0)}\equiv \omega_{BL},\, \vev{(1,1,3,0)}\equiv \omega_{R} ,\, \vev{(1,1,3,+2)}\equiv \sigma,
\ee
where $\omega_{BL,R}$ are real and $\sigma$ can be made real by
a phase redefinition of the $126_H$. Different VEV configurations
trigger the spontaneous breakdown of the $SO(10)$ symmetry into several qualitatively distinct subgroups. Namely, for $\sigma= 0$ one finds (in an obvious notation)
\begin{align}
\label{vacua}
&\omega_{R}= 0,\, \omega_{BL}\neq 0\; : & 3_c\, 2_L\, 2_R\, 1_{BL}\,, \nn \\[0.5ex]
&\omega_{R}\neq 0,\, \omega_{BL}= 0\; : & 4_{C} 2_L 1_R\,, \nn \\[0.5ex]
&\omega_{R}\neq 0,\, \omega_{BL}\neq 0\; : & 3_c\, 2_L\, 1_R\, 1_{BL}\,,\\[0.5ex] 
&\omega_{R}=-\omega_{BL}\neq 0\; : & \mbox{flipped}\, 5'\, 1_{Z'}\,, \nn \\[0.5ex]
&\omega_{R}=\omega_{BL}\neq 0\; :  & \mbox{standard}\, 5\, 1_{Z}\,, \nn
\end{align}
with  $5\, 1_{Z}$ and $5'\, 1_{Z'}$ standing for the two inequivalent
embeddings of the SM hypercharge operator $Y$ into $SU(5) \otimes U(1)\subset SO(10)$ usually called the ``standard'' and the ``flipped'' $SU(5)$ scenarios~\cite{DeRujula:1980qc,Barr:1981qv}, respectively. 
In the standard case, 
$
Y=T^{3}_R+\tfrac{1}{2}T_{BL}
$
belongs to the $SU(5)$ algebra and the orthogonal Cartan generator
$Z$ is given by
$
Z =-4T^{3}_R+3T_{BL}
$.
In the flipped ($5'1_{Z'}$) case, the right-handed isospin
assignment of quarks and leptons 
is turned over so that the flipped hypercharge generator reads
$Y'=-T^{3}_R+\tfrac{1}{2}T_{BL}$.
Accordingly, the additional $U(1)_{Z'}$ generator reads
$Z' =4T^{3}_R+3T_{BL}$ (for further details see, e.g., ref.~\cite{Bertolini:2009es}).

For $\sigma \neq 0$ all the intermediate gauge symmetries (\ref{vacua}) are spontaneously broken down to the SM group, with the exception of the last case which maintains the $SU(5)$ subgroup unbroken.

%===========================================================
\subsection{Gauge unification and the scalar octet mass}
\label{sect:unification}
%===========================================================
There are several basic criteria we impose on each vacuum of the theory featuring a light color octet scalar. Besides the very consistency of the gauge unification picture at two loops we demand  compatibility with the proton lifetime constraints and require a reasonably high seesaw scale in order to support a renormalizable (and, hence, potentially predictive) implementation of the seesaw mechanism.
This program for the minimal $SO(10)$ setting here considered has been initiated in~\cite{Bertolini:2012im}. The goal of the current study is to include the relevant two-loop running effects and to assess their impact on the breaking pattern and the scalar spectrum. 

%===========================================================
\subsubsection{Two-loop gauge unification constraints}
%===========================================================
The Higgs setting we are considering here (i.e., with an adjoint Higgs driving the $SO(10)$ breaking)
was shown to provide phenomenologically interesting breaking patterns only if at least 
the one-loop effective potential is considered~\cite{Bertolini:2009es}. Gauge unification constraints
and the shape of the scalar spectrum have been discussed in~\cite{Bertolini:2012im}, where
it is shown that an acceptable $B-L$ scale for the renormalizable seesaw implementation is obtained only if the scalar spectrum exhibits "light" colored scalar states (in particular, an octet or a sextet).
This result improved on previous analyses (see for instance \cite{Bertolini:2009qj} and 
refs.~\cite{Chang:1984qr,Buccella:1989za,Deshpande:1992au} for earlier studies) based on the 
minimal survival hypothesis (MSH) \cite{del Aguila:1980at,Mohapatra:1982aq}, 
where just the scalar spectrum necessary for the spontaneous symmetry breaking at each stage is assumed at the corresponding scale. This allows a for systematic analysis, albeit preliminary to a detailed study of the vacuum constraints. 

In what follows we shall consider only the setting with the light color octet scalar transforming as a weak doublet, i.e., $H_{8}\equiv (8,2,+\tfrac{1}{2})$,
for its potential relevance to the TeV physics scale. A discussion of the shape of the relevant one-loop scalar spectrum, the effects of the heavy scalar thresholds 
and the constraints obtained from the gauge unification and the absolute neutrino mass scale will be given
in Section~\ref{results}.

%===========================================================
\paragraph{The tree-level scalar spectrum.\label{treescalarspectrum}}
Adopting the convention in which the mass term in the Lagrangian is written as $\tfrac{1}{2}\psi^T M^2 \psi$, where $\psi=(\phi, \Sigma^\ast, \Sigma)$ is a 297-dimensional vector, 
the scalar spectrum is obtained readily by evaluating the relevant functional scalar mass matrix of the schematic form
\be
M^2(\phi, \Sigma^*, \Sigma)=\left(
\begin{array}{lll}
V_{\phi\phi} & V_{\phi\Sigma^\ast} & V_{\phi\Sigma} \\
V_{\Sigma^\ast\phi} & V_{\Sigma^\ast\Sigma^\ast} & V_{\Sigma^\ast\Sigma} \\
V_{\Sigma\phi} & V_{\Sigma\Sigma^\ast} & V_{\Sigma\Sigma}
\end{array}
\right)
\label{M2matrix}
\ee
 on the SM vacuum.
The subscripts here denote the derivatives of the scalar potential with respect to a specific set of fields.
Subsequently, this matrix can brought to a block-diagonal form (i.e., into the SM basis) by a unitary transformation.
The complete tree-level spectrum is found in Appendix~B of ref.~\cite{Bertolini:2012im}.

%===========================================================
\paragraph{One-loop scalar spectrum.} Conceptually, a complete two-loop analysis of the unification pattern requires a thorough understanding of the one-loop spectrum of the theory.
This, however, is extremely demanding in full generality. On the other hand, for the sake of the current analysis it is sufficient to focus on the most prominent one-loop corrections to the scalar masses, namely, those that cure the instabilities of the tree level potential~\cite{Bertolini:2009es}.

The leading loop-induced nonlogarithmic correction in the scalar sector comes from tadpoles~\cite{Aoki:1982ed} which, among other contributions, yield the ``universal" ($SO(10)$ symmetric) shift to the scalar masses via the $\tau$ term in the potential. It is not difficult to see that the only source of a $\tau^{2}$-proportional nonlogarithmic term is associated to the renormalization of the stationarity conditions imposed on the scalar potential. Diagrammatically, it corresponds to a special cluster of one-loop graphs contributing to the one-point function of $45_{H}$, cf.~\cite{Bertolini:2012im}. 
Given the $SO(10)$ structure of the relevant $\tau$-vertex in~(\ref{V45126}) one finds
\be\label{leadingoneloopcorrection}
\Delta M^{2}_{\tau^{2}}=\frac{35\tau^{2}}{32\pi^{2}}\, 
\ee 
that can be viewed as a typical one-loop correction of the scalar masses in the vicinity of the GUT
scale. This is numerically relevant namely for those states in $45_H$ that are tachyonic at the tree-level~\cite{Bertolini:2012im}, thus allowing for a simplified treatment of the scalar spectrum.

%===========================================================
\paragraph{Effective gauge theories and matching scales:\label{sect:scheme}} 
The generic structure of the vacua supporting a light color octet scalar, 
namely $\omega_{R}\gg (\omega_{BL},\sigma$), was discussed in~\cite{Bertolini:2012im}. This suggests that the breaking of the $SO(10)$ 
gauge  symmetry can be conveniently described by the following series of effective gauge
theories~\cite{Weinberg:1980wa}
\bea
&SO(10) &\label{scheme} \\ &\downarrow^{\mu_{2}} & \nn\\ 
&SU(4)_{C}\otimes SU(2)_{L}\otimes U(1)_{R}& \nn\\ 
&\downarrow^{\mu_{1}} & \nn\\ 
& SU(3)_{c}\otimes SU(2)_{L}\otimes U(1)_{Y}& \quad {\rm (SM+H_8)}\nn\\
&\downarrow^{\mu_{0}} & \nn\\ 
& SU(3)_{c}\otimes SU(2)_{L}\otimes U(1)_{Y}& \nn \quad {\rm(SM)}
\eea
where $\mu_{2}>\mu_{1}>\mu_{0}$ denote the relevant matching scales. Unlike for the first two steps where the gauge symmetry of the effective theory changes (as does the associated vector boson spectrum with all subtleties associated to that), the third arrow corresponds to the decoupling of just the light $H_8$ at a certain scale $\mu_{0}$ with no change in the gauge sector and, as such, it is almost trivial. In the numerical simulation in Sect.~\ref{results} the scales $\mu_{1,2}$ are naturally chosen in the vicinity of the masses of the gauge bosons associated to the two relevant symmetry breakings triggered by the VEVs $\omega_{R}$ and $\sigma$, i.e., 
\be\label{naturalmus}
\mu_{2}=g^{0}_{G}\omega_{R}\,,\;\mu_{1}=g^{0}_{G}\sigma\,, 
\ee 
where $g_{G}^{0}= 0.6$ is the typical value of the gauge couplings at (or around) the GUT scale identified in~\cite{Bertolini:2012im}.

Strictly speaking, the scheme (\ref{scheme}) is naturally implied for $\sigma\gtrsim \omega_{BL}$ which, indeed, corresponds to our main interest in settings with a  maximum allowed $\sigma$, cf.~Section~\ref{sect:seesaw_scale}. In the opposite case, i.e., for $\omega_{BL}\gtrsim \sigma$, one may for simplicity consider an extra stage with an intermediate $SU(3)_{c}\otimes SU(2)_{L}\otimes U(1)_{R}\otimes U(1)_{BL}$ symmetry. Since, however, this is mostly a matter of language (as the difference between the two approaches can be essentially subsumed into a set of extra threshold effects in the simple scheme above) we shall consider only four effective theories along with three effective matching scales for all settings.  
The interested reader can find further remarks on the consistency of this approach in section~\ref{results:two_loops}.

%===========================================================
\paragraph{The two-loop beta functions:\label{betafunctions}} 
In what follows we shall pass through all the steps in (\ref{scheme}) and list all the $a_{i}$ and $b_{ij}$ coefficients entering the 
two-loop renormalization group equations for the gauge couplings~\cite{Jones:1974mm,Caswell:1974gg,Jones:1981we,Machacek:1983tz}\footnote{The two-loop contribution of the Yukawa couplings is here neglected.
As a matter of fact we checked that the top-Yukawa coupling contributes to less than the $10\%$ to the difference between the one- and two-loop results. In addition, such a contribution
produces an almost uniform shift on the evolved gauge couplings, thus affecting only the determination of the unified gauge coupling  (cf.~e.g.~the discussion in Sect.~IV D of \cite{Bertolini:2009qj}).}
\be
\frac{{\rm d}}{{\rm d}t}\alpha^{-1}_{i}=-a_{i}-\frac{b_{ij}}{4\pi}\alpha_{j}
\label{alpha2loops}\,,
\ee
where 
\be\label{tdef}
t=\frac{1}{2\pi}\log \mu/M_{Z}
\ee  
provided
\bea
a_{i}&=&- \frac{11}{3} C_2(G_i) + \frac{4}{3}\kappa S_2(F_i) + \frac{1}{3} \eta S_2(S_i)\,, \\
b_{ij}&=& 
\left[- \frac{34}{3} \left( C_2(G_i) \right)^2  \right.        \label{Gp2loops} \\ 
&+&                \left( 4 C_2(F_i) + \frac{20}{3} C_2(G_i) \right) \kappa S_2(F_i) \nn \\
&+&         \left. \left( 4 C_2(S_i) + \frac{2}{3} C_2(G_i) \right) \eta S_2(S_i) \right]\delta_{ij} \nn \\
&+&        4 \Big[  \kappa C_2(F_j) S_2(F_i)
                         + \eta C_2(S_j) S_2(S_i)  \Big]  \nn
\eea
(no summation over $i$). Here $S_{2}$ and $C_{2}$ denote the index (including multiplicity factors) and the quadratic Casimir of a given representation, $\kappa=1,\frac{1}{2}$ for Dirac and Weyl fermions and
$\eta=1, \frac{1}{2}$ for complex and real scalar fields, respectively.
For a detailed account of the subtleties related to the case with more than a single abelian gauge factor the interested reader may refer to the 
discussion in ref.~\cite{Bertolini:2009qj} and references therein. 
For $|\omega_{j}(t-t_{0})| < 1$, with $\omega_j \equiv a_j \alpha_j(t_0)$, the expression
\be 
\alpha_i^{-1}(t) - \alpha_i^{-1}(t_{0}) = - {a_i}\ (t-t_{0}) + \frac{b_{ij}}{4\pi a_{j}}
                                     \log\left[1- \omega_j( t-t_{0})\right]\,
\ee
is, to a great accuracy, a solution of  \eq{alpha2loops}.

Above $\mu_{2}$ the spectrum of the theory under consideration is defined by the 45-dimensional adjoint representation containing the gauge fields, three copies of the 16-dimensional matter spinors, a real 45-dimensional adjoint scalar and a complex 126-dimensional antiselfdual component of the 5-index antisymmetric $SO(10)$ tensor. This altogether yields
\be
a=-37/3\,,\quad b=9529/6\,.
\ee   

The scale $\mu_2$  characterizes the $SO(10)$ breaking to $SU(4)_C \otimes SU(2)_L \otimes U(1)_R$. At this stage, the propagating gauge bosons fill the $(15,1,0)\oplus(1,3,0)\oplus(1,1,0)$ reducible representation of the gauge group, together with matter multiplets in three copies of the $(4,2,0)\oplus (\overline{4},1,+\tfrac{1}{2})\oplus (\overline{4},1,-\tfrac{1}{2})$ representation and complex scalars in $(\overline{10},1,-1)\oplus (15,2,+\tfrac{1}{2})$. Note that the latter is just the minimal set of fields that may trigger the subsequent steps of the gauge symmetry breaking and, thus, the natural expectation of the minimal survival shape of the scalar spectrum is conformed. With all this at hand one has
\be
a_{i}=(-7,-\tfrac{5}{6},\tfrac{59}{6})\,,\quad
b_{ij}=
\renewcommand{\arraystretch}{1.25}\left(\begin{array}{ccc}
 \tfrac{265}{2}& \tfrac{57}{2} &  \tfrac{43}{2}\\
  \tfrac{285}{2}& \tfrac{217}{6} &  \tfrac{15}{2}\\
 \tfrac{645}{2} & \tfrac{45}{2} &  \tfrac{101}{2}
\end{array}\right)\,,
\ee
where $i,j\in\left\{4_{C},2_{L},1_{R}\right\}$.

Below $\mu_1$ the effective theory is the SM plus $H_8$ scalar. The gauge fields are grouped into $(8,1,0)\oplus(1,3,0)\oplus(1,1,0)$ while the matter resides in three copies of $(3,2,\tfrac{1}{6})\oplus (1,2,-\tfrac{1}{2})\oplus (\overline{3},1,-\tfrac{2}{3})\oplus (\overline{3},1,+\tfrac{1}{3})\oplus (1,1,-1)$. The surviving $H_8$ scalars transform as $(8,2,+\tfrac{1}{2})\oplus (1,2,+\tfrac{1}{2})$. 
The RGE beta coefficients read:
\be
a_{i}=(-5,-\tfrac{11}{6},\tfrac{49}{10})\,,\quad
b_{ij}=
\renewcommand{\arraystretch}{1.25}\left(\begin{array}{ccc}
58& \tfrac{45}{2} &  \tfrac{47}{10}\\
  60& \tfrac{139}{6} &  \tfrac{33}{10}\\
 \tfrac{188}{5} & \tfrac{99}{10} &  \tfrac{271}{50}
\end{array}\right)\,,
\ee
where $i,j\in\left\{3_{c},2_{L},1_{Y}\right\}$.

After $H_8$ is integrated out (at $\mu_0\equiv M_8$), the minimal set of the SM fields (including one Higgs doublet) yields:
\be
a_{i}=(-7,-\tfrac{19}{6},\tfrac{41}{10})\,,\quad
b_{ij}=
\renewcommand{\arraystretch}{1.25}\left(\begin{array}{ccc}
-26& \tfrac{9}{2} &  \tfrac{11}{10}\\
  12& \tfrac{35}{6} &  \tfrac{9}{10}\\
 \tfrac{44}{5} & \tfrac{27}{10} &  \tfrac{199}{50}
\end{array}\right)\, .
\ee
At $M_{Z}$ (i.e., at $t=0$) the $SU(5)$-normalized SM gauge couplings are required to fall into the current 1-$\sigma$ bands
\begin{align}
\alpha_1&= 0.0169225 \pm 0.0000039 \, ,\nn \\
\label{alphaMZ}
\alpha_2 &= 0.033735 \pm 0.000020\, , \\
\alpha_3 &= 0.1173 \pm 0.0007 \nn\, , 
\end{align}
were these data refer to the modified minimal subtraction scheme ($\overline{\rm{MS}}$) 
in the full SM, i.e.~the top being not integrated out \cite{Mihaila:2012pz,Martens:2010nm,Beringer:1900zz}.

%===========================================================
\paragraph{Threshold corrections.\label{sect:thresholdeffects}} 

What remains to be detailed is the matching between all the gauge theories in (\ref{scheme}). 
The general form of the one-loop matching condition between effective theories in the framework of
dimensional regularization has been discussed in \cite{Weinberg:1980wa,Hall:1980kf}.
Let us consider first a simple gauge group $G$ spontaneously broken into subgroups $G_i$.
The one-loop matching for the gauge couplings can be then written as
\be
g_{i}^{-2} = g^{-2} - \lambda_i(\mu)\,,
\label{2LmatchGUT}
\ee
where 
\bea\label{lambda0}
\lambda_i(\mu)&=&\frac{1}{48\pi^{2}} \left[ \Tr T_{iV}^{2} - 21\Tr T_{iV}^{2}\log\frac{M_{V}}{\mu}
\right. \\
&+&\left.   8\Tr T_{iF}^{2}\log\frac{M_{F}}{\mu}+\Tr T_{iS}^{2}\log\frac{M_{S}}{\mu}\right]\;\;\;\nn
\eea
with $V$, $F$ and $S$ denoting the massive vectors, fermions and scalars that are integrated out at the matching scale $\mu$ (classified with respect to the  preserved symmetries~$G_i$)\footnote{As a matter of fact, one may choose any proper subgroup $G'_i$ of $G_i$ for the classification of the decoupled fields. This freedom is ensured by the identity $S_{2}(R^{G})=\sum_{R^{G'}} S_{2}(R^{G'})$.}. Let us also note that  $T_{i}$ stand for the corresponding group generators and the relevant multiplicities are looked after the traces. 

In the notation of \eq{Gp2loops} the matching corrections can be written as 
\bea\label{lambda}
\lambda_i(\mu)&=&\frac{1}{48\pi^{2}}S_{2}(V_{i})+\frac{1}{8\pi^{2}}\left[-\frac{11}{3}\Tr T_{V_{i}}^{2}\log\frac{M_{V_{i}}}{\mu}\right. \\
&+&\left.\frac{4}{3}\kappa\Tr T_{F_{i}}^{2}\log\frac{M_{F_{i}}}{\mu}+\frac{1}{3}\eta\Tr T_{S_{i}}^{2}\log\frac{M_{S_{i}}}{\mu}\right]\; , \nn
\eea
where the (Feynman gauge) Goldstone bosons have been conveniently included in the scalar part of the expression. Note that $S_{2}(V_{i})$ can here be written as $C_2(G)-C_2(G_i)$.

When multiple $U(1)$ factors are present one must be careful about the abelian mixing effects; in what follows we
shall follow the notation and discussion of ref.~\cite{Bertolini:2009qj}.
Consider the breaking of $N$ copies of $U(1)$ gauge factors to a subset of $M$ elements $U(1)$ (with $M<N$).
Denoting by $T_n$ ($n=1,...,N$) and by $\widetilde T_m$ ($m=1,...,M$)
their properly normalized generators we have
\be
\widetilde T_m = P_{mn} T_n
\label{U1charges}
\ee
with the orthogonality condition $P_{mn}P_{m'n} = \delta_{mm'}$. Let us denote
by $g_{na}$ ($n,a=1,...,N$) and by $\widetilde g_{mb}$ ($m,b=1,...,M$) the matrices
of abelian gauge couplings above and below the breaking scale respectively.
Writing the abelian gauge boson mass matrix in the broken vacuum and  identifying
the massless states one finds the following matching condition
\be
(\widetilde g \widetilde g^T )^{-1} = P \left(g g^T \right)_{\rm eff}^{-1} P^T\,,
\label{2LU1match}
\ee
where
\be
g_{\rm eff\ AB}^{-2} \equiv  g_{\rm AB}^{-2} - \lambda_{\rm AB}(\mu)\, ,
\label{geff}
\ee
with $A,B = 1,...,N$.
Equation (\ref{2LU1match}) depends on the choice of basis for the $U(1)$ charges (via $P$)
but it is invariant under orthogonal rotations of the gauge boson fields ($gO^TOg^T=gg^T$).
Notice that whenever the decoupled states are classified by multiple $U(1)$ charges
the presence of nonvanishing off-diagonal entries in $\lambda_{\rm AB}$ is crucial for
the correct matching even if no multiple-$U(1)$-symmetric stage is actually considered (i.e., for diagonal
$g_{\rm AB}^{-2}$).

The general case of a gauge group $ U(1)_{1} \otimes ... \otimes U(1)_{N}\otimes G_1\otimes ... \otimes G_{N'}$
spontaneously broken to $U(1)_{1} \otimes ... \otimes U(1)_{M}$ with $M\leq N+N'$
is taken care of by replacing $(gg^T)^{-1}$  in \eq{2LU1match}
with the block-diagonal $(N+N')\times (N+N')$ matrix
\be
(G G^T)^{-1} = \mbox{Diag}\left[(g g^T)^{-1},g_i^{-2}\right] \ ,
\label{2Lmatchgen}
\ee
thus providing, together with the extended version of \eq{U1charges}, the necessary generalization of \eq{2LmatchGUT}.

%\subparagraph{Decoupling at $\mu_{2}$:} 
 At the $SO(10)\to SU(4)_{C}\otimes SU(2)_{L}\otimes U(1)_{R}$ breaking scale $\mu_{2}$
the following components are decoupled (the subscripts denote the origin of the listed multiplets in terms of the 
$SO(10)$ scalar irreps [cf.~also TABLEs \ref{tab:45decomp} and \ref{tab:126decomp}], while the superscripts V and GB indicate their vector and/or Goldstone-boson nature): $(1,1,\pm 1)^{V}$, $(1,1,\pm 1)^{\rm GB}_{45}$, $(6,2,\pm \tfrac{1}{2})^{V}$, $(6,2,\pm \tfrac{1}{2})^{\rm GB}_{45}$, 
$(6,1,0)_{126}$, $(10,3,0)_{126}+\rm{h.c.}$, $(\overline{10},1,0)_{126}+\rm{h.c.}$, $(\overline{10},1,+1)_{126}+\rm{h.c.}$, 
$(1,1,0)_{45}$, $(1,3,0)_{45}$, $(15,1,0)_{45}$, and the heavy eigenstate of the 
$(15,2,\pm\tfrac{1}{2})_{126}+\rm{h.c.}$ system.  
Let us stress that the last item of the list above corresponds to the state orthogonal to the component hosting the light SM doublet Higgs that has been  identified as ``active'' throughout the 421 stage in Sect.~\ref{betafunctions}.

%\subparagraph{Decoupling  at $\mu_{1}$:} 
At the $SU(4)_{C}\otimes SU(2)_{L}\otimes U(1)_{R}\to {\rm SM}+H_8$  matching scale, we have to take into account that the SM hypercharge generator is a weighted average of that of the  $U(1)_{BL}$ subgroup of $SU(4)_{C}$ and the $U(1)_{R}$ charge. In terms of the canonically normalized abelian charges one has  $Y=\sqrt{\tfrac{3}{5}} T_{R}^{3}+\sqrt{\tfrac{2}{5}} X$, where $X=\sqrt{\tfrac{3}{8}}(B-L)$ obeys $\Tr X^{2}=1$ in the defining 10-dimensional vector representation of $SO(10)$. In what follows, it will be useful to define the hypercharge projector
$
P_Y = \left(\sqrt{\tfrac{3}{5}},\ \sqrt{\tfrac{2}{5}} \right)
\label{P_Y}
$ for which $Y=P_{Y}(T_{R}^{3},X)^{T}$.

In terms of the $3_{c}2_{L}1_{R}1_{BL}$ quantum numbers, the fields that decouple at $\mu_{1}$
are (with subscripts and superscripts as above): $(\overline{3},1,0,-\tfrac{4}{3})^{\rm V}+\rm{h.c.}$, the light eigenstate of the $(\overline{3},1,-1,+\tfrac{2}{3})_{126}\oplus (\overline{3},1,0,-\tfrac{4}{3})_{45}+\rm{h.c.}$ system playing the role of the associated Goldstone boson\footnote{Even though the specific shape of the $\lambda$ matrix (\ref{lambdamatrix}) does depend on the details of the projection of the light eigenstate onto the two underlying components $(\overline{3},1,-1,+\tfrac{2}{3})_{126}\oplus (\overline{3},1,0,-\tfrac{4}{3})_{45}$ there is actually no need to worry about this because both theses components individually provide the same contribution to the $P_{Y}\lambda(\mu_{1})P_{Y}^{T}$ factor in (\ref{Ymatching}) and so does any of their properly normalised linear combinations.}, $(1,1,-1,+2)_{126}+\rm{h.c.}$, $(\overline{6},1,-1,-\tfrac{2}{3})_{126}$, $(3,2,+\tfrac{1}{2}, +\tfrac{4}{3})_{126}+\rm{h.c.}$ and $(3,2,-\tfrac{1}{2}, +\tfrac{4}{3})_{126}+\rm{h.c.}$. 
An explicit example of the one-loop massive spectrum can be found in \Table{sample82p12}.
With all this at hand one can construct the corresponding $2\times 2$ abelian matching matrix,
\be
\label{lambdamatrix}\lambda(\mu_{1})=
\left(\begin{array}{cc}
 \lambda_{RR}&  \lambda_{RX}\\
\lambda_{XR}&  \lambda_{XX}
\end{array}\right)\,,
\ee
which subsequently enters the hypercharge matching condition
\be\label{Ymatching}
\alpha^{-1}_{Y}(\mu_{1})=4\pi P_{Y}\left[(GG^{T})^{-1}(\mu_{1})-\lambda(\mu_{1})\right]P_{Y}^{T}\ .
\ee
In particular,
\be 4\pi (GG^{T})^{-1}(\mu_{1}) = 
\left(\begin{array}{cc}
 \alpha^{-1}_{R}(\mu_{1})&  0\\
0&  \alpha^{-1}_{4}(\mu_{1})
\end{array}\right)\,
\ee
is the matrix of the abelian gauge couplings which, in the case of our interest, is trivially diagonal since there is no explicit intermediate 3211 running in the relevant sequence of the effective gauge theories (\ref{scheme}). A specific example of the $\lambda(\mu_{1})$ matrix is reported in Appendix~\ref{AppB}.

%\subparagraph{Decoupling at $\mu_{0}$:} 
Finally, at $\mu_{0}$ fixed to the mass of $H_8$ (henceforth denoted by $M_8$) the effective theory becomes the pure SM. Needless to say, since the matching scale is conveniently chosen at the mass of the decoupled state, the matching is technically trivial.

%================================================================
\subsubsection{\label{sect:protonlimits}Proton lifetime limits}
%================================================================
%..................................................
\paragraph{Gauge-induced $d=6$ proton decay:\label{d6protondecay}}
%..................................................

With a detailed information on the heavy spectrum of the model at hand one can rather accurately estimate the proton lifetime for all specific settings of interest. The master formula adjusted for the $SO(10)$ gauge content reads~\cite{Dorsner:2009mq,Nath:2006ut}
\bea
& & \Gamma({p\to \pi^{0}e^{+}}) = \frac{\pi}{4}A_{L}^{2}(1+F+D)^{2}\frac{|\alpha|^{2}}{f_{\pi}^{2}}m_{p}\alpha_{G}^{2}\times\nn \\
&  & \times \left[A_{SR}^{2}\left(\frac{1}{M^{2}_{(X,Y)}}+\frac{1}{M^{2}_{(X'\!,Y')}}\right)^{2}+\frac{4 A_{SL}^{2}}{M^{4}_{(X,Y)}}\right]\label{protonwidth}\,,
\eea
where $A_{L}=1.25$ is the factor encoding the renormalization from the electroweak scale to the proton mass, $D=0.81$, $F=0.44$,  $\alpha=-0.011~{\rm GeV}^{3}$ and $f_{\pi}=139$~MeV are phenomenological parameters given by the chiral perturbation theory and lattice, $m_{p}=938.3$~MeV is the proton mass and $\alpha_{G}$ is the GUT-scale gauge coupling. Let us also note that $A_{SL}$ and $A_{SR}$ are the renormalization factors of the relevant $d=6$ proton decay operators from the GUT scale to the weak scale,
\be\label{ASLSR}
A_{SL(R)}=\prod_{i=1}^{3}\prod_{P}^{M_{Z}\leq m_{P}<M_{G}}\left[\frac{\alpha_{i}(m_{P+1})}{\alpha_{i}(m_{P})}\right]^{\frac{\gamma_{L(R)i}}{\sum_{Q}^{M_{Z}\leq m_{Q}\leq m_{P}}\Delta a_{iQ}}}.
\ee
Here, $\gamma_{L}=(\tfrac{23}{20},\tfrac{9}{4},2) $,  $\gamma_{R}=(\tfrac{11}{20},\tfrac{9}{4},2) $,  $P$ and $Q$ are labels of states relevant at each stage and $\Delta a_{iQ}$ is the contribution of the field $Q$ to the one-loop beta function; cf. Sect.~\ref{betafunctions}.
Furthermore, $M_{(X,Y)}$ and $M_{(X'\!,Y')}$ denote the masses of the underlying GUT-scale gauge bosons (i.e., those transforming, respectively, as 
$(\overline{3},2,+\tfrac{5}{6})$ and $(\overline{3},2,-\tfrac{1}{6})$ under the SM gauge group). 
The latter are given, for the model under consideration, in Appendix C of ref.~\cite{Bertolini:2012im}. 

Due to the lack of information about the flavor sector of the theory, we have set all the flavor matrices governing the baryon and lepton number violating currents~\cite{FileviezPerez:2004hn,Dorsner:2004xa} coupled to the heavy gauge fields to a $3\times 3$ unit matrix; hence, we do not entertain any accidental cancellations in the relevant amplitudes which, in turn, makes our results conservative.   

Given this, we shall implement namely the basic constraint corresponding to the current best Super-Kamiokande (SK) 
limit~\cite{:2012rv}:
\be\label{limit:SK2012}
\tau(p\to e^{+}\pi^{0})_{\rm SK, 2012}> 1.3 \times 10^{34}\, {\rm years}\,,
\ee
together with a pair of speculative Hyper-Kamiokande (HK) limits that are assumed to be reached by 2025 and 2040, respectively~\cite{Abe:2011ts}:
\bea
\label{limit:HK2025}\tau(p\to e^{+}\pi^{0})_{\rm HK, 2025}& > &  9 \times 10^{34}\, {\rm years}\,,\\
\label{limit:HK2040}\tau(p\to e^{+}\pi^{0})_{\rm HK, 2040} & > &  2 \times 10^{35}\, {\rm years}\,.
\eea
In all figures of the next section, points obeying these three limits shall be, consecutively, denoted by light grey, dark grey and black colors. 
\vskip 2mm
%..................................................
\paragraph{Scalar-induced $d=6$ proton decay:\label{scalard6protondecay}}
%..................................................
In general, the scalar-induced $d=6$ proton decay operators are expected to be suppressed with respect to the gauge-driven ones due to the extra flavor factors associated to the first generation Yukawa couplings in the relevant baryon-number-violating currents. In the framework under consideration this expectation is further justified by the fact that the potentially dangerous colored triplets ($\Delta_{c}$) never fall far from the GUT scale and, thus, never really compete with the gauge sector. To be on the safe side, we shall follow the strategy defined in the previous one-loop analysis~\cite{Bertolini:2012im} 
and implement a conservative lower bound of $m_{\Delta_{c}}\gtrsim 10^{14}$~GeV.
\vskip 2mm
%..................................................
\paragraph{$d>6$ induced proton decay:\label{d7protondecay}}
%..................................................
Due to the rather specific shape of the scalar spectrum and, in particular, the absence of baryon number violation in the $H_8$ couplings, the $d>6$ proton decay is expected to be highly suppressed with respect to the $d=6$ type of transitions. For further comments, the interested reader is deferred to Sect. III of reference~\cite{Bertolini:2012im}.   

%================================================================
\subsubsection{Absolute neutrino mass scale\label{sect:seesaw_scale}}
%================================================================
Another phenomenological issue we shall consider concerns the seesaw  mechanism, which, for a potentially predictive scheme, is implemented at the renormalizable level via the VEVs of the relevant RH and LH triplets from $126_{H}$. Assuming that the associated Yukawa coupling of $126_{H}$ is of order 1, the natural size of the seesaw scale (i.e., the VEV of the RH triplet) is in the $10^{13}$~GeV ballpark (and larger for smaller $126_{H}$ Yukawa couplings). Thus, in what follows, we show our results for two conservative regions corresponding to $\sigma \gtrsim 10^{12}$ GeV and $\sigma \gtrsim 10^{13}$ GeV, respectively. On the other hand, as we shall see, these limits do not affect the absolute upper bound for $M_{8}$ in any substantial manner, cf.~Fig.~\ref{octettwoloopseesawlimits}.

%===========================================================
\subsection{Results\label{results}}
%===========================================================
The discussion of the results of the numerical simulation shall be divided into two parts. First, we shall discuss the shape of the  parameter space that remains open when the two-loop gauge unification and the proton lifetime constraints in Eqs. (\ref{limit:SK2012})--(\ref{limit:HK2040}) are taken into account. Secondly, we shall discuss the general upper bounds on $M_8$ and their possible changes when, e.g., additional requirements related to the absolute neutrino mass scale in renormalizable implementations of the seesaw are imposed, cf.~Sect.~\ref{sect:seesaw_scale}. Finally, we shall briefly comment on the robustness of the results with respect to the
variation of the intermediate matching scales, the possible effects of an additional $10_{H}$ in the Higgs sector and other relevant sources of uncertainties.

%================================================================
\subsubsection{Consistent two-loop gauge unification\label{results:two_loops}}
\begin{figure}[t]
\includegraphics[width=7.5cm]{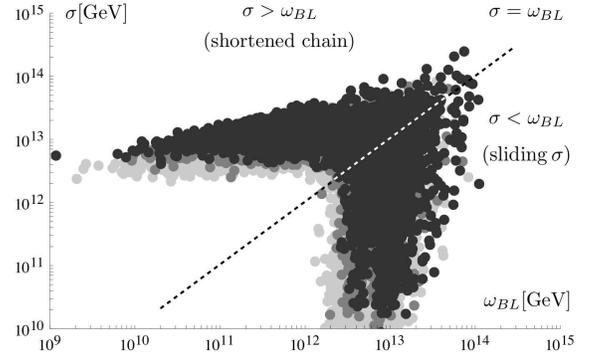}
\caption{\label{plotparameters} The $|\omega_{BL}|-\sigma$ plane depicting the allowed part of the parameter space that supports a consistent two-loop gauge unification with a color octet scalar in the TeV domain (color code defined in Sect.~\ref{d6protondecay}). For efficiency reasons, we reject all points that, in the one-loop approximation, yield $\sigma$ below $10^{10}$ GeV; this leads to a reduction of the plot density in the uninteresting lowest $\sigma$ region at two loops.} 
\end{figure}

%================================================================
The shape of the relevant part of the parameter space corresponding to a stable vacuum supporting consistent gauge unification patterns with a light $H_{8}$, compatible with the current proton lifetime limits, is best seen from the $\omega_{BL}-\sigma$ plane depicted in Fig.~\ref{plotparameters}. The two regimes associated to the points above and below the $\omega_{BL}=\sigma$ diagonal correspond to the ``shortened'' $SO(10)\to 4_{C}2_{L}1_{R}\to {\rm SM+H_8}\to {\rm SM}$ symmetry-breaking chain (upper-left part of Fig.~\ref{plotparameters}) and to the ``sliding $\sigma$ regime'' underpinning the $SO(10)\to 4_{C}2_{L}1_{R}\to 3_{c}2_{L}1_{R}1_{BL}\to {\rm SM+H_8}\to {\rm SM}$ symmetry-breaking pattern  (lower-right part of Fig.~\ref{plotparameters}), respectively.
 The fact that both these regions can stretch so far from the $\omega_{BL}\sim\sigma$ diagonal has to do with the fact that in neither of the two cases the subdominant VEV plays any significant role. This is quite clear for $\omega_{BL}<\sigma$  because in such a situation none of the scalar masses is governed by $\omega_{BL}$ while in the opposite case (i.e., for $\sigma<\omega_{BL}$) the lower VEV governs only SM-singlet (scalar) states which do not affect the evolution of the SM gauge couplings.
 
Let us also notice that the difference between the latter case and the breaking pattern~(\ref{scheme}) based on the two matching scales $\mu_{1,2}$ is just formal because the short hierarchy between the heavier states driven by $\omega_{BL}$ and  those that decouple at $\sigma$ is  well accounted for by the threshold corrections in the matching equation~(\ref{Ymatching}).
%#####

%================================================================
\subsubsection{Color octet scalar mass bounds\label{generallimits}}
\begin{figure}[t]
\includegraphics[width=7.5cm]{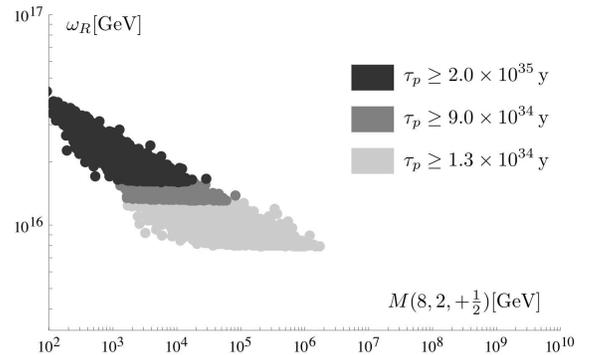}
\caption{\label{octettwoloopgeneric} The $H_8$ mass range allowed by the two-loop unification constraints and matter stability bounds. The present day bound on matter stability sets an upper bound on the $H_8$ mass of about 2000 TeV. The blurry lower-left boundary of the allowed region is an artefact of the numerical procedure in which, for efficiency reasons, we dump all uninteresting points with  $\sigma$ below $10^{10}$ GeV at one-loop. On the other hand, the upper-right boundary (which is the one that sets the stringent limit on the mass of $H_{8}$) corresponding to the sharp upper cut in Fig.~\ref{plotparameters} is enforced by the unification constraints and, as such, it is a robust feature of the model.}
\end{figure}

%================================================================
%\paragraph{The general case.}
Turning our attention to the allowed range for the mass of the light colored octet scalar $H_8$, the most general situation is depicted in Fig.~\ref{octettwoloopgeneric}.~We see that the two-loop effects, as well as the improved implementation of the proton lifetime limits shrink the formerly identified range (see  
Fig.~6 in 
ref.~\cite{Bertolini:2012im}) by about four orders of magnitude. Hence, the overall consistency of the minimal $SO(10)$ GUT, together with the present day bounds on the matter stability, requires $H_8$ to be lighter  than about $2000$~TeV. Remarkably enough, this limit gets further reduced to about 20 TeV if no proton decay is detected  up to a time scale of $2 \times 10^{35}$~years corresponding to the maximum foreseen Hyper-Kamiokande reach, 
cf.~\eq{limit:HK2040}. 
%#####
%#####

\begin{figure}[t]
\includegraphics[width=8.8cm,height=4.2cm]{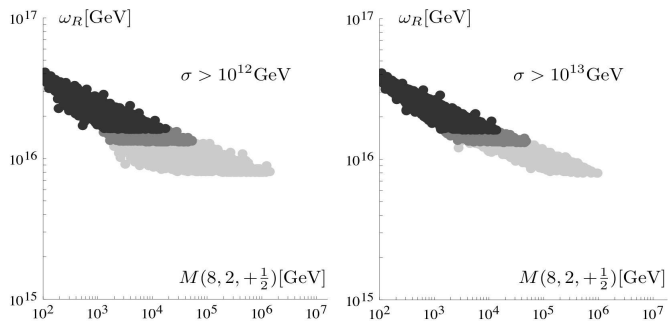}
\caption{\label{octettwoloopseesawlimits} The same as  in Fig.~\ref{octettwoloopgeneric} with the inclusion of the $B-L$ scale limits specified in Sect.~\ref{sect:seesaw_scale} (namely, $\sigma \gtrsim 10^{12}$ GeV on the left and $\sigma \gtrsim 10^{13}$ GeV on the right). Since these constraints affect the bottom-left 
parts of the allowed regions, they preserve the upper limits quoted in Sect.~\ref{generallimits}. In all cases $H_8$ turns out to be lighter than about 2000~TeV. The bound shrinks to about 20 TeV for a proton lifetime above $10^{35}$~years (black area).}
\end{figure}

Note also that these bounds are robust with respect to the cuts imposed on the $B-L$ scale $\sigma$, as discussed in Sect.~\ref{sect:seesaw_scale}. By looking at Fig.~\ref{octettwoloopseesawlimits} we see that increasing $\sigma$ does not affect the color octet mass upper limit (i.e., the rightmost points of a given color); the reason is that these points lay in the uppermost part of Fig.~\ref{plotparameters} and, as such, they are the last ones to  be affected by additional constraints on $\sigma$. Nevertheless,  the whole allowed region is eventually wiped out for $\sigma \gtrsim 2\times 10^{14}$~GeV which sets a hard upper limit on the seesaw scale in this scenario.
%##### 
%#####

\subsubsection{A specific example}
\begin{figure}[h]
\includegraphics[width=8cm,height=5cm]{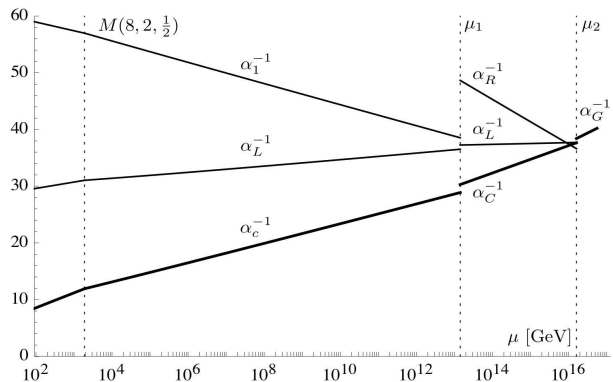}
\caption{\label{samplesolution}A sample two-loop gauge unification pattern corresponding to one of the allowed solutions in Figs.~\ref{plotparameters}, \ref{octettwoloopgeneric} and \ref{octettwoloopseesawlimits}, see TABLE~\ref{TableSampleParameters} and \ref{sample82p12} for further details. The gauge running proceeds through four stages  corresponding from top down to the $SO(10)$, $SU(4)_C \otimes SU(2)_L \otimes U(1)_R$, the SM+$H_8$ and, finally, the pure SM settings. The visible discontinuities in the non-Abelian gauge couplings at the upper two matching points are due to the large threshold corrections generated by the states decoupled at each scale.  } 
\end{figure}

For the sake of illustration, let us detail a specific solution corresponding to one of the black points in Figs. \ref{plotparameters}-\ref{octettwoloopseesawlimits} (i.e., those with the proton lifetime exceeding $2\times 10^{35}$~years). 
The parameter-space coordinates of this sample point are given in \Table{TableSampleParameters}, the full breakdown of the bosonic part of the corresponding spectrum is given in \Table{sample82p12} in Appendix~\ref{AppSolution} and, finally, the gauge unification pattern is depicted in Fig.~\ref{samplesolution}. 

%#####
%#####

\begin{table}[h]
\begin{tabular}{c|c}
\hline\hline
Parameter   & Value \\
\hline
$\omega_{R}$  & $1.62 \times 10^{16}$ {\rm GeV}   \\
$\omega_{BL}$ & $5.41  \times 10^{11}$  {\rm GeV}   \\
$\sigma$  &  $-1.43  \times 10^{13}$ {\rm GeV}  \\
$a_{0}$ &  $ 0.74 $   \\
$\alpha$ &  $ -0.53 $   \\
$\beta_{4} $ &  $ 0.65 $ \\
$\beta_{4}'$ &  $ -0.47 $  \\
$\gamma_{2}$ &  $ 0.09 $  \\
$\lambda_{0}$ &  $ -0.36 $  \\
$\lambda_{2}$ &  $ 0.33  $ \\
$\lambda_{4}$ &  $ 0.62  $\\
$\lambda_{4}' $ &  $ 0.41  $\\
\hline
$M(8,2,+\tfrac{1}{2})$  & $1.90\times 10^{3}$ {\rm GeV} \\
\hline\hline
\end{tabular}
\caption{\label{TableSampleParameters}Parameters underpinning the sample spectrum given in \Table{sample82p12} in Appendix~\ref{AppSolution} with the corresponding unification pattern depicted in Fig.~\ref{samplesolution}. The value of the $\tau$ parameter can be obtained from the requirement that the color octet scalar has the mass specified in the last row.}
\end{table}

\subsubsection{Further comments and consistency checks}

\paragraph{Robustness of the two-loop renormalization group analysis.}
In order to verify the robustness of our numerical results we checked that, for a given set of the GUT-scale initial conditions, the results of the two-loop renormalization group analysis are only marginally depending on the particular choice of the matching scales $\mu_{1,2}$ in the vicinity of the main decoupling thresholds. As an example, by varying $\mu_{i}$'s by as much as a factor of 3 leads to just minuscule variations in the low-energy values of $\alpha_{i}$'s; by defining $\Delta_{i}\equiv\alpha_{i}^{-1}(M_{Z})_{\rm perturbed}-\alpha_{i}^{-1}(M_{Z})_{\rm reference}$, where the latter corresponds to the ``standard'' choice of $\mu_{i}$'s as in Eq~(\ref{naturalmus}), we obtain $|\Delta_{1}-\Delta_{2}|\lesssim 0.01$ and $|\Delta_{2}-\Delta_{3}|\lesssim 0.1$.\footnote{These numbers can be compared for instance with the low-energy uncertainties in 
\eq{alphaMZ}, yielding $\Delta(\alpha_1^{-1}) \approx 0.014$, $\Delta(\alpha_2^{-1}) \approx 0.018$ and 
$\Delta(\alpha_3^{-1}) \approx 0.051$.}

\paragraph{Effect of an extra $10_H$ in the Higgs sector.\label{MSH}} 

Concerning the impact of the so far neglected extra 10-dimensional scalar representations that represent a minimal extra measure in order to arrange a potentially viable flavor pattern (cf. Sect.~\ref{FCNC})  there is actually no need to undertake the technically demanding task of adding $10_{H}$ into the scalar potential (\ref{scalpotgen}) and redoing the same analysis from scratch. 
It is clear that its colored triplet components (cf.~\Table{tab:10decomp}) mix heavily with the other colored triplets in the potential and, thus, they live near to or at the GUT scale. Hence, the leading RGE effect of an additional $10_{H}$ comes from the need to mix its doublet  component $(1,2,\tfrac{1}{2})_{10}$ with $(15,2,\tfrac{1}{2})_{126}$ in order to provide the right SM flavor structure. Such a mixing can be significant if and only if both these multiplets survive down to about the 421 breaking scale and, hence, the extra $(1,2,\tfrac{1}{2})_{10}$ should be considered among the active states in the relevant part of Sect.~\ref{betafunctions}.

In order to gauge the effect of such an extra  $(1,2,\tfrac{1}{2})_{10}$,  we can take advantage of the results
obtained in the simple framework based on the MSH. In this approach, one ignores the details of the scalar spectrum and assumes a simple clustering of the minimal set of scalars needed for the spontaneous symmetry breaking at each given scale. Having subtracted the ``noise" of the detailed spread of the scalar spectrum, this approach allows us to assess (rather reliably) the effects of  a given extra state at a desired stage of the gauge running. 
Such a simplified analysis shows that, for a fixed $H_8$ mass, the net effect of an extra $(1,2,\tfrac{1}{2})$ running throughout the $4_{C}2_{L}1_{R}$ stage is a further decrease of the GUT scale, $M_{G}$, while increasing the seesaw scale, $M_{B-L}$. Quantitatively, for a $\mathcal{O}(\rm{TeV})$ $H_8$ 
the inclusion of a weak doublet between $M_{G}$ and $M_{B-L}$, 
results into a $-10 \%$ and a $+75 \%$ shift in $M_{G}$ and in $M_{B-L}$, respectively. 

The fact that for a fixed $M_8$ the GUT scale (proportional to $\omega_{R}$) 
is slightly lowered makes the proton lifetime constraints more severe. 
Thus, with the extra doublet at play, the general upper limit on $M_8$ derived in the previous sections (see namely Fig.~\ref{octettwoloopgeneric}) can only get stronger. 
At the same time, the seesaw scale may be increased up to a factor two, but with essentially no effects on the light $H_8$ mass constraints (cf.~Fig.~\ref{octettwoloopseesawlimits}). 
Considering a more realistic case with the doublet components of $10_H$ lying somewhere between $\mu_2$
and $\mu_1$ the net effect of the extra $10_{H}$ is further weakened.

\paragraph{$M_8-M_G$ correlation.}

The $M_8 - M_G$ correlation in Fig.~\ref{octettwoloopgeneric} can be qualitatively understood  by noticing that:
(i) the $H_8$ contribution to the one-loop beta functions is such that $a_3 > a_2 > a_1$ (cf.~\Table{sample82p12});
(ii) taking the variation of the matching condition at $\mu_{1}$, 
$\alpha_1^{-1} - \tfrac{2}{5} \alpha_{C}^{-1} = \tfrac{3}{5} \alpha_{R}^{-1}$ (cf.~also \fig{samplesolution}), with respect to $M_8$ 
there is a partial cancellation on the left-hand side so that $\partial \alpha_{R}^{-1} / \partial M_{8} \approx 0$. 
Thus, the value of the GUT scale depends predominantly on the convergence point of $\alpha_{L}$ and $\alpha_{C}$. As a consequence, the fact that higher 
unification scales select lighter $H_8$ is just a geometrical consequence of $a_3 > a_2$. 
For a similar discussion of the correlation between the GUT scale and the mass of a light  $(8,2,\tfrac{1}{2})$  octet in $SU(5)$ models see, e.g.,~\cite{Dorsner:2006dj,Dorsner:2007fy,Perez:2007rm}.

\paragraph{$\Delta\alpha_3(M_Z)$ uncertainty.} 

Among the most prominent sources of systematic errors one should certainly quote the 
$\mathcal{O}(0.6\%)$ uncertainty in $\alpha_3 (M_Z)$ [cf.~\eq{alphaMZ}], which translates 
into a conservative $\mathcal{O}(10\%)$ uncertainty in the determination 
of the seesaw and the GUT scales. The latter is, in turn, responsible for a larger $\mathcal{O}(50\%)$ 
uncertainty in the upper bound on the $H_8$ mass, which is mainly due to 
the almost flat slope of the $M_8-M_{G}$ mass correlation (cf.~\fig{octettwoloopgeneric}).

\paragraph{Planck-scale physics.} 

Worth of mention is the general fragility of grand unifications with respect to the Planck-scale ($M_{Pl}$) effects which, 
given the proximity of the GUT scale and $M_{Pl}$, may not be entirely negligible.
Concerning their possible impact on, e.g., the proton lifetime estimates, the most important of effect is the Planck-scale induced 
violation of the canonical normalization of the SM gauge fields~\cite{Calmet:2008df,Chakrabortty:2008zk} 
due to the higher-order corrections to the gauge kinetic term emerging already at the $d=5$ 
level: ${\cal L}^{(5)}\ni {\rm Tr} [ F_{\mu\nu}HF^{\mu\nu}]/M_{Pl}$.
Here $H$ is any scalar in the theory which can couple to a pair of adjoint representations, 
i.e., any field appearing in the symmetric part of the decomposition of their tensor product. 
For a GUT-scale VEV of $H$, this induces a percent-level effect which, 
after a suitable redefinition of the gauge fields, leads to similar-size shifts in the GUT-scale matching conditions which, on the other hand, are comparable 
to the two-loop effects. 

Remarkably, if the GUT-scale symmetry breaking in $SO(10)$ is triggered by the VEV of $45_{H}$, this problem is absent because 
${\rm Tr} [F_{\mu\nu}45_{H}F^{\mu\nu}]=0$ due to the fact that 45 is not in the symmetric part of the $45\otimes 45$ decomposition [recall that $(45\otimes 45)_{\rm sym}=54\oplus 210\oplus 770$]. Thus, the minimal $SO(10)$ scheme with the adjoint-driven Higgs mechanism is uniquely robust with respect to this class of quantum gravity effects (see also the discussion in ref.~\cite{Bertolini:2012be}). 
This makes the symmetry-breaking analysis more reliable and, hence, 
admits in principle for a strong reduction of this type of theoretical uncertainties in the proton lifetime estimates.

%%%%%%%%%%%%%%%%%%%%%%%%%%%%%%%%%%%%%%%%%%%
\section{Flavour and electroweak observables}
%%%%%%%%%%%%%%%%%%%%%%%%%%%%%%%%%%%%%%%%%%%
\label{FlavourandEW}

The low-energy phenomenology of a light $\mathcal{O}(\rm{TeV})$-scale color octet scalar 
which couples directly to the quarks and the Higgs boson 
is severely constrained by the flavor 
and electroweak precision observables. 
In this section we describe such constraints from the point of view of the minimal $SO(10)$ GUT.

%================================================================
\subsection{\label{FCNC} Tree-level FCNC}
%================================================================

Color octet scalars with the same electroweak quantum numbers as the SM Higgs doublet 
have rather special properties since the suppression 
of the FCNC could naturally follow from the group structure and representation contents of the theory \cite{Manohar:2006ga}. 
One possible way to achieve this is, indeed, imposing the minimal flavor violation (MFV) 
ansatz\footnote{The effective field theory approach to MFV consists in the assumptions that: 
(i) the full effective field theory is formally invariant under the $SU(3)^5$ flavor symmetry of the SM,
and (ii) the SM Yukawas, which are promoted to spurion fields transforming under the $SU(3)^5$ symmetry,  
are the only irreducible sources of flavor breaking. 
Under these assumptions, only the scalars with the SM quantum numbers 
$(1,2,+\tfrac{1}{2})$ and $(8,2,+\tfrac{1}{2})$ can couple to the quarks \cite{Manohar:2006ga}
and the amount of flavor violation beyond the SM is controlled by the standard Yukawas \cite{Gresham:2007ri}.} 
\citelist{Chivukula:1987py}{D'Ambrosio:2002ex}.
On the other hand, the $SO(10)$ symmetry provides nontrivial 
constraints on the flavor sector of the effective 
SM$+ H_8$ theory. 
As we shall see, the analysis of the relevant structure reveals 
the existence of an intrinsic flavor-protection mechanism even 
without any additional ``horizontal'' assumption like MFV. 

In order to proceed further we need to specify the $SO(10)$ Yukawa sector. 
At this stage, we shall consider two minimal, albeit rather different, realizations of the 
Yukawa sector which can potentially reproduce the pattern of fermion masses 
and mixings.\footnote{A complete and conclusive analysis on the viability of 
these settings in the $45_H \oplus 126_H$ model is still missing. 
For studies partially addressing the problem see e.g.~\cite{Babu:1992ia,Bajc:2005zf,Joshipura:2011nn}.}
Adding a complex\footnote{The most minimal option of a real $10_H$ (enforcing $v_u^{10} = v_d^{*10}$) 
is not phenomenologically viable since it predicts a ``wrong'' GUT-scale mass relation 
$m_t \approx m_b$ \cite{Bajc:2005zf}.} $10_H$ representation 
into the Yukawa sector and focussing only on the mass sum-rules for the up- and down-quark mass matrices 
we have the following:
\begin{itemize}
\item Minimal field content \cite{Bertolini:2012im},
\begin{align}
\label{MUmfc}
M_U &= Y_{10}' v_d^{*10} \, , \\
\label{MDmfc}
M_D &= Y_{10} v_d^{10} + Y_{126} v_d^{126} \, ,
\end{align}
\item Peccei-Quinn (PQ) symmetry \cite{Bajc:2005zf},
\begin{align}
\label{MUPQ}
M_U &= Y_{10} v_u^{10} + Y_{126} v_u^{126}  \, , \\
\label{MDPQ}
M_D &= Y_{10} v_d^{10} + Y_{126} v_d^{126} \, ,
\end{align}
\end{itemize}
where $Y_{10}$ and $Y_{126}$ are the usual $SO(10)$ Yukawa couplings  while $Y_{10}'$ is an extra Yukawa available in the nonsupersymmetric models with a complex $10_{H}$. 
The presence of $v_{u,d}$ in (\ref{MUPQ})-(\ref{MDPQ}) indicates that there are two light Higgs doublets in the low-energy effective theory in the PQ case while there is just one Higgs doublet in the minimal scenario (\ref{MUmfc})-(\ref{MDmfc}).
Note also that all the Yukawa couplings above are symmetric in the flavor space due to the gauge structure of the relevant $SO(10)$ invariants. 

The second option is related to the implementation of the PQ solution to the strong CP problem
in $SO(10)$.
The need for an invisible axion requires an additional Higgs
representation in the potential \cite{Lazarides:1981kz,Holman:1982tb,Mohapatra:1982tc}, minimally a $16_H$. In spite of that the PQ symmetry actually reduces the number of couplings in the scalar potential and in the Yukawa sector with respect to the first case. In addition to that, a viable dark matter (DM) candidate, the axion, is 
available~\cite{Kim:2008hd}. It is worth recalling that a DM candidate can be devised in the ``minimal'' setting as well, albeit requiring an extreme fine-tuning~\cite{Nemevsek:2012cd}.

Returning to the issue of the induced FCNCs, the interactions of $H_8$ with the quarks are dictated by the $Y_{126}$ matrix, namely 
\begin{multline}
\label{H8quarkint}
Y_{126} 16_M 16_M 126^*_H + \rm{h.c.} \\
\ni Y_{126} \left( q u^c H_{8u} + q d^c H_{8d} \right) + \rm{h.c.} \\
\ni Y_{126} \left( q u^c c_\theta H_8 + q d^c s_\theta H_8^* \right) + \rm{h.c.}
\, .
\end{multline}
Natural flavor conservation in the neutral currents then  
%i.e.~no tree-level FCNC, 
requires $Y_{126}$ to be 
(almost) diagonal in the same basis as $M^{\rm{diag}}_U$ and $M^{\rm{diag}}_D$. 

Remarkably enough, in the PQ case [cf.~\eqs{MUPQ}{MDPQ}]
the FCNCs turn out to be $V_{\rm CKM}$-suppressed due to the special flavor structure of the relevant Yukawa couplings.  
This can be intuitively understood as follows: $SO(10)$  
enforces symmetric Yukawa matrices and hence $V_{u_L} = V_{u_R}$ and $V_{d_L} = V_{d_R}$ where $V_{u_{L,R}}$ and $V_{d_{L,R}}$ denote 
the biunitary transformations which diagonalize the up- and down-quark mass matrices, respectively. 
Then, any flavor violation in the quark sector is encoded in the misalignment between $V_{u_{L}}$ and $V_{d_{L}}$ and, thus, the $H_8$-mediated FCNCs are controlled by the CKM mixing matrix as in the MFV setting~\cite{Gresham:2007ri}. 

It is perhaps also worth noting that in the minimal case [cf.~\eqs{MUmfc}{MDmfc}] 
the FCNCs are entirely absent if~$\left[ Y_{10}, Y_{126} \right] = 0$. 
The latter condition is not phenomenologically acceptable
in the PQ setting since it would imply a diagonal CKM mixing matrix.  

It is also important to stress that while $Y_{126}$ is correlated to fermion masses and mixings, 
the angle $\theta$ (cf.~\eq{H8quarkint}) is a function of a few scalar potential parameters [Eq.~(B11) in \cite{Bertolini:2012im}]. 
Thus, the couplings of $H_8$ to the fermions are quite constrained and the compatibility of a very light octet scalar with the flavor and electroweak (nonoblique) observables \cite{Manohar:2006ga,Gresham:2007ri}, 
as, e.g.,~$K^0-\overline{K}^0$ mixing, $B \rightarrow X_s \gamma$ and $Z \rightarrow b \overline{b}$, 
must be ultimately checked. 

Needless to say, the PQ setting requires a detailed study of fermion masses and mixings, up to date missing to our knowledge.  In particular, one must keep in mind that the orthogonality
of the hypercharge and the PQ currents further constraints  the
VEVs of the light $H_{u,d}$ doublets by enforcing $v_u^2=v_d^2$.

%================================================================
\subsection{\label{CS} Custodial symmetry}
%================================================================

The most crucial among the electroweak precision tests are those related to the breaking of the custodial symmetry\footnote{Large contributions to the oblique parameter $S$ can be easily avoided even for a sub-TeV-scale $H_8$, 
while retaining at the same time a sizable impact on the SM $H \rightarrow \gamma\gamma$ rate \cite{Manohar:2006ga}.}. 
Indeed, the Higgs VEV induces a tree-level mass splitting between the charged $H^+_8$ and neutral $H^{R,I}_8$ (real and imaginary) components of $H_8$ which determines a nonvanishing 
contribution to the $T$ parameter~\cite{Manohar:2006ga} 
%\begin{multline}
%\Delta T \approx 
%\frac{\Delta m^2_{+,R} \Delta m^2_{+,I}}{6 \pi \sin^2\theta_W M_W^2 m^2_{+,R,I}} \\
%\approx 3.55 \times 10^{-5} \frac{\Delta m^2_{+,R} \Delta m^2_{+,I}}{m^2_{+,R,I}} \ \rm{GeV}^{-2}  \, ,
%\end{multline}
\be
\frac{\Delta m^2_{+,R} \Delta m^2_{+,I}}{m^2_W m^2_{+,R,I}}  \approx 0.23\  \Delta T \, ,
\label{DeltaT}
\ee
where $\Delta m^2_{1,2} \equiv m^2_1 - m^2_2$ and $m^2_{+,R,I}$ is the smallest among the masses. The approximate \eq{DeltaT} holds with high accuracy even for $m_{+,R,I}$ as light as
a few hundred GeV with $\Delta m$'s  at the 100 GeV level. As a matter of fact the present experimental
fits lead to $T = 0.02 \pm 0.11$~\cite{Erler:2012wz}; requiring a 1-$\sigma$ saturation of $T$ and taking
$\Delta m^2_{+,R} \approx \Delta m^2_{+,I}$
one obtains $\Delta m < 140$ GeV for the lightest mass of $300$ GeV and
 $\Delta m < 250$ GeV for the lightest mass of $1$ TeV.
Since the mass splitting between the charged and neutral components is
at most of the order of the electroweak scale such constraints can be naturally satisfied. 

Large contributions to the $T$ parameter can be also avoided when custodial symmetry limits in the parameter space of the 
Higgs doublet and $H_8$ effective scalar potential is considered\footnote{See refs.~\cite{Manohar:2006ga,Burgess:2009wm} 
for a detailed discussion of precision electroweak 
constraints on the scalar octet interactions.},
albeit this generally requires fine-tuned relations among the scalar couplings.
Though the embedding into $SO(10)$ could make such a limit nontrivial, a quantitative answer 
requires an extension of the $45_H \oplus 126_H$ scalar potential analyzed in \cite{Bertolini:2012im}, 
including those representations which are needed for a realistic description of 
fermion masses and mixings and have a nonzero projection on the 
Higgs doublets, like, e.g.,~a (complex) $10_H$. This, however, is beyond the scope of the present work.

Finally, let us comment on the current experimental limits on the $H_8$ mass obtained namely from the direct LHC searches. 
While the recent experimental analysis based on dijet pair signatures \cite{ATLAS:2012pu,CMS:2012yf} 
exclude $H_8$ masses up to about $2$ 
TeV, the low-energy window $200-320$ GeV is not yet ruled out due to 
a gap in sensitivity to the soft jets \cite{Altmannshofer:2012ur}. On the other hand, the most recent four-jet final-state studies of the ATLAS~\cite{ATLAS:2012ds} and CMS~\cite{Chatrchyan:2013izb} collaborations
seem to exclude also this range.

%%%%%%%%%%%%%%%%%%%%%%%%%%%%%%%%%%%%%%%%%%%
\section{Conclusions}
%%%%%%%%%%%%%%%%%%%%%%%%%%%%%%%%%%%%%%%%%%%

As emphasised recently a light (sub-TeV) colored octet scalar may help in explaining the $H\to\gamma\gamma$ anomaly still present in the LHC data. In a preceding paper we  scrutinized the role of intermediate-scale colored scalar states in the minimal $SO(10)$ grand unification. The presence of such states together with the unification constraints and the associated large threshold effects allow for the scale of the $B-L$ breaking in the desired ballpark for the neutrino mass generation via standard seesaw. Moreover, the model may be prone to future experimental testability due to a relatively rapid proton decay inherent to this class of scenarios\footnote{Another viable threshold configuration identified in~\cite{Bertolini:2012im} 
which allows for a large enough seesaw scale is that of an intermediate-scale color sextet scalar $(6,3,+\tfrac{1}{3})$. 
Preliminary results indicate that two-loop effects further lower the GUT scale thus reducing the physically allowed domain of this class of solutions \cite{workinprog}. 
}. 

In this paper, we provide a significantly refined study of the setup with a light color octet scalar by including the leading two-loop gauge running effects together with a more detailed analysis of the proton decay width. Focusing on the interesting anticorrelation between the proton lifetime and the mass of the light $H_8$ we find that the present data on the matter stability require $M_{8}$ below about $2000$ TeV; yet stronger limits at the level of few tens of TeV are expected if proton decay is not observed even at the next generation facilities. In all cases a sizeable fraction of the parameter space allows for $H_8$ masses within the reach of the LHC.

%%%%%%%%%%%%%%%%%%%%%%%%%%%%%%%%%%%%%%%%%%%
\subsection*{Acknowledgments}
%%%%%%%%%%%%%%%%%%%%%%%%%%%%%%%%%%%%%%%%%%%

S.~B. is partially supported by the italian MIUR Grant No.~2010YJ2NYW\_001 and by the EU Marie Curie ITN UNILHC, Grant Agreement PITN-GA-2009-237920. The work of L.~D.~L. is supported by the DFG through the SFB/TR 
9 ``Computational Particle Physics.'' 
In the initial phase, the work of M.~M. was supported by the Marie Curie Intra European Fellowship
within the 7th European Community Framework Programme
FP7-PEOPLE-2009-IEF, Contract No. PIEF-GA-2009-253119;  by the EU
Network Grant No. UNILHC PITN-GA-2009-237920; by the Spanish MICINN
Grants No.  FPA2008-00319/FPA and No. MULTIDARK CAD2009-00064
(Consolider-Ingenio 2010 Programme); and by the Generalitat
Valenciana Grant No. Prometeo/2009/091. In the later stages, M.~M. was supported by the Marie Curie Career Integration Grant within the 7th European Community Framework Programme
FP7-PEOPLE-2011-CIG, Contract No. PCIG10-GA-2011-303565; and by the Research Proposal No. MSM0021620859 of the Ministry of Education, Youth and Sports of the Czech Republic.

\appendix

\section{A sample scalar spectrum\label{AppSolution}}
\begin{table}[h]
\begin{tabular}{c|c|c|c|c}
\hline\hline
Multiplet $Q$ & \; Type\;  & Eigenstate & $\Delta a_{Q}$ & Mass [GeV]  \\
\hline
$\bf (8,2,+\tfrac{1}{2})$ & {\bf CS} & 1 & $(2, \tfrac{4}{3}, \tfrac{4}{5})$ & $\bf 1.9 \times 10^{3}$ \\
\hline
$(\overline{3},1,-\tfrac{2}{3})$ & VB & 1 & $(-\tfrac{11}{6}, 0, -\tfrac{44}{15})$ & $1.2 \times 10^{13}$ \\
\hline
$(3,1,+\tfrac{2}{3})$ & VB & 1 & $(-\tfrac{11}{6}, 0, -\tfrac{44}{15})$ & $1.2 \times 10^{13}$ \\
\hline
$(\overline{3},1,-\tfrac{2}{3})$ & GB & 1 & $(\tfrac{1}{6}, 0, \tfrac{4}{15})$ & $1.2 \times 10^{13}$ \\
\hline
$(1,1,0)$ & VB & 1 & $(0, 0, 0)$ & $2.7 \times 10^{13}$ \\
\hline
$(1,1,0)$ & GB & 1 & $(0, 0, 0)$ & $2.7 \times 10^{13}$ \\
\hline
$(3,2,+\tfrac{1}{6})$ & CS & 2 & $(\tfrac{1}{3}, \tfrac{1}{2}, \tfrac{1}{30})$ & $8.2 \times 10^{13}$ \\
\hline
$(3,2,+\tfrac{7}{6})$ & CS & 1 & $(\tfrac{1}{3}, \tfrac{1}{2}, \tfrac{49}{30})$ & $1.1 \times 10^{14}$ \\
\hline
$(1,2,+\tfrac{1}{2})$ & {RS} & 1 & $(0, \tfrac{1}{12}, \tfrac{1}{20})$ & $1.1 \times 10^{14}$ \\
\hline
$(1,1,0)$ & RS & 2 & $(0, 0, 0)$ & $4.2 \times 10^{15}$ \\
\hline
$(\overline{3},1,-\tfrac{2}{3})$ & CS & 2 & $(\tfrac{1}{6}, 0, \tfrac{4}{15})$ & $4.2 \times 10^{15}$ \\
\hline
$(\overline{6},1,-\tfrac{4}{3})$ & CS & 1 & $(\tfrac{5}{6}, 0, \tfrac{32}{15})$ & $5.4 \times 10^{15}$ \\
\hline
$(1,1,0)$ & RS & 3 & $(0, 0, 0)$ & $5.4 \times 10^{15}$ \\
\hline
$(8,1,0)$ & RS & 1 & $(\tfrac{1}{2}, 0, 0)$ & $6.2\times 10^{15}$ \\
\hline
$(6,3,+\tfrac{1}{3})$ & CS & 1 & $(\tfrac{5}{2}, 4, \tfrac{2}{5})$ & $7.4 \times 10^{15}$ \\
\hline
$(3,3,-\tfrac{1}{3})$ & CS & 1 & $(\tfrac{1}{2}, 2, \tfrac{1}{5})$ & $7.4 \times 10^{15}$ \\
\hline
$(1,3,-1)$ & CS & 1 & $(0, \tfrac{2}{3}, \tfrac{3}{5})$ & $7.4 \times 10^{15}$ \\
\hline
$(1,3,0)$ & RS & 1 & $(0, \tfrac{1}{3}, 0)$ & $8.4 \times 10^{15}$ \\
\hline
$\bf(\overline{3},2,+\tfrac{5}{6})$ & \bf VB & 1 & $(-\tfrac{11}{3}, -\tfrac{11}{2}, -\tfrac{55}{6})$ & $\bf 9.7 \times 10^{15}$ \\
\hline
$\bf(3,2,-\tfrac{5}{6})$ & \bf VB & 1 & $(-\tfrac{11}{3}, -\tfrac{11}{2}, -\tfrac{55}{6})$ & $\bf 9.7 \times 10^{15}$ \\
\hline
$\bf (3,2,-\tfrac{5}{6})$ & \bf GB & 1 & $(\tfrac{1}{3}, \tfrac{1}{2}, \tfrac{5}{6})$ & $\bf 9.7 \times 10^{15}$ \\
\hline
$\bf(\overline{3},2,-\tfrac{1}{6})$ & \bf VB & 1 & $(-\tfrac{11}{3}, -\tfrac{11}{2}, -\tfrac{11}{30})$ & $\bf 9.7 \times 10^{15}$ \\
\hline
$\bf(3,2,+\tfrac{1}{6})$ & \bf VB & 1 & $(-\tfrac{11}{3}, -\tfrac{11}{2}, -\tfrac{11}{30})$ & $\bf 9.7 \times 10^{15}$ \\
\hline
$\bf (3,2,+\tfrac{1}{6})$ & \bf GB & 1 & $(\tfrac{1}{3}, \tfrac{1}{2}, \tfrac{1}{30})$ & $\bf 9.7 \times 10^{15}$ \\
\hline
$(\overline{3},1,+\tfrac{1}{3})$ & CS & 1 & $(\tfrac{1}{6}, 0, \tfrac{1}{15})$ & $1.2 \times 10^{16}$ \\
\hline
$(\overline{3},1,+\tfrac{1}{3})$ & CS & 2 & $(\tfrac{1}{6}, 0, \tfrac{1}{15})$ & $1.8 \times 10^{16}$ \\
\hline
$(1,1,-1)$ & VB & 1 & $(0, 0, -\tfrac{11}{5})$ & $1.9 \times 10^{16}$ \\
\hline
$(1,1,+1)$ & VB & 1 & $(0, 0, -\tfrac{11}{5})$ & $1.9 \times 10^{16}$ \\
\hline
$(1,1,+1)$ & GB & 1 & $(0, 0, \tfrac{1}{5})$ & $1.9 \times 10^{16}$ \\
\hline
$(1,1,+1)$ & CS & 2 & $(0, 0, \tfrac{1}{5})$ & $2.0\times 10^{16}$ \\
\hline
$(\overline{3},1,+\tfrac{1}{3})$ & CS & 3 & $(\tfrac{1}{6}, 0, \tfrac{1}{15})$ & $2.0 \times 10^{16}$ \\
\hline
$(\overline{6},1,-\tfrac{1}{3})$ & CS & 1 & $(\tfrac{5}{6}, 0, \tfrac{2}{15})$ & $2.0 \times 10^{16}$ \\
\hline
$(3,2,+\tfrac{7}{6})$ & CS & 2 & $(\tfrac{1}{3}, \tfrac{1}{2}, \tfrac{49}{30})$ & $2.3 \times 10^{16}$ \\
\hline
$(1,2,+\tfrac{1}{2})$ &{RS} & 2 & $(0, \tfrac{1}{12}, \tfrac{1}{20})$ & $2.3 \times 10^{16}$ \\
\hline
$(8,2,+\tfrac{1}{2})$ & CS & 2 & $(2, \tfrac{4}{3}, \tfrac{4}{5})$ & $2.3 \times 10^{16}$ \\
\hline
$(3,2,+\tfrac{1}{6})$ & CS & 3 & $(\tfrac{1}{3}, \tfrac{1}{2}, \tfrac{1}{30})$ & $2.3 \times 10^{16}$ \\
\hline
$(1,1,+2)$ & CS & 1 & $(0, 0, \tfrac{4}{5})$ & $3.3 \times 10^{16}$ \\
\hline
$(\overline{3},1,+\tfrac{4}{3})$ & CS & 1 & $(\tfrac{1}{6}, 0, \tfrac{16}{15})$ & $3.3 \times 10^{16}$ \\
\hline
$(\overline{6},1,+\tfrac{2}{3})$ & CS & 1 & $(\tfrac{5}{6}, 0, \tfrac{8}{15})$ & $3.3 \times 10^{16}$ \\
\hline
$(1,1,0)$ & RS & 4 & $(0, 0, 0)$ & $5.6 \times 10^{16}$ \\
\hline\hline
\end{tabular}
\caption{\label{sample82p12}
A sample spectrum featuring a light $(8,2,+\tfrac{1}{2})$ multiplet. 
The relevant scalar potential parameters are given 
in~\Table{TableSampleParameters}. $\Delta a_{Q}$ indicate 
the shifts in the one-loop beta-function entering formula~(\ref{protonwidth}) due to a given multiplet $Q$ . The light threshold and the vector bosons defining the GUT scale are in boldface.
As a consistency check,  $a_{SM}+\sum \Delta a_{Q}=(-\tfrac{37}{3},-\tfrac{37}{3},-\tfrac{37}{3})$.
}
\end{table}
Details of the scalar spectrum corresponding to the sample two-loop solution displayed in Fig.~\ref{samplesolution} are given in \Table{sample82p12}. For the sake of brevity, for each multiplet $Q$ therein we present only its contribution to the one-loop part of the gauge beta-function ($\Delta a_{Q}$) which is enough to reconstruct the $A_{SL,SR}$ factors~(\ref{ASLSR}) governing the one-loop evolution of the effective $d=6$ proton decay operators and, hence, the proton partial width (\ref{protonwidth}); in the sample case one has $\Gamma({p\to \pi^{0}e^{+}})\approx (2.0\times 10^{35} {\rm years})^{-1}$. In \Table{sample82p12}, the acronyms CS, RS, VB, GB stay for complex scalars, real scalars, vector bosons and would-be Goldstone bosons, respectively. %The eigenstates are numbered from lighter to heavier.

\vfill\eject

\section{One-loop matching\label{AppB}}
For completeness, we report the detailed form of the structures entering the matching of the gauge couplings at the $SO(10)-4_C 2_L 1_R$ threshold $\mu_{2}$ and at the $4_C 2_L 1_R-{\rm SM}+H_8$ threshold $\mu_{1}$, respectively. 
\vskip 2mm
%==================================
\paragraph*{The $SO(10)\to 4_{C}2_{L}1_{R}$ matching at $\mu_{2}$:}
\bea
& & \lambda_{C}=\nn \\
& & \frac{1}{48 \pi^{2}}\left[
4
-44\log\frac{M(3,2,+\tfrac{1}{6})_{\rm VB}}{\mu_{2}}
+2\log\frac{M(3,2,+\tfrac{1}{6})_{\rm GB}}{\mu_{2}}
\right.
\nn\\
& & \qquad
-44\log\frac{M(3,2,-\tfrac{5}{6})_{\rm VB}}{\mu_{2}}
+2\log\frac{M(3,2,-\tfrac{5}{6})_{\rm GB}}{\mu_{2}}
\nn\\
& & \qquad
+\log\frac{M(3,1,+\tfrac{2}{3})_{\rm CS}^{(2)}}{\mu_{2}}
+\log\frac{M(3,1,-\tfrac{4}{3})_{\rm CS}}{\mu_{2}}
\nn\\
& & \qquad
+2\log\frac{M(3,2,+\tfrac{1}{6})_{\rm CS}^{(2)}}{\mu_{2}}
+2\log\frac{M(3,2,+\tfrac{7}{6})_{\rm CS}^{(2)}}{\mu_{2}}
\nn\\
& & \qquad
+\log\frac{M(3,1,+\tfrac{1}{3})_{\rm CS}^{(1)}}{\mu_{2}}
+\log\frac{M(3,1,+\tfrac{1}{3})_{\rm CS}^{(2)}}{\mu_{2}}
\nn\\
& & \qquad
+\log\frac{M(3,1,+\tfrac{1}{3})_{\rm CS}^{(3)}}{\mu_{2}}
+3\log\frac{M(3,3,+\tfrac{1}{3})_{\rm CS}}{\mu_{2}}
\nn\\
& & \qquad
+5\log\frac{M(6,1,+\tfrac{1}{3})_{\rm CS}}{\mu_{2}}
+5\log\frac{M(6,1,-\tfrac{2}{3})_{\rm CS}}{\mu_{2}}
\nn\\
& & \qquad
+15\log\frac{M(\overline{6},3,-\tfrac{1}{3})_{\rm CS}}{\mu_{2}}
+3\log\frac{M(8,1,0)_{\rm RS}}{\mu_{2}}
\nn\\
& & \qquad\left.
+12\log\frac{M(8,2,+\tfrac{1}{2})_{\rm CS}^{(2)}}{\mu_{2}}\right] \, ,
\nn
\eea
\bea
& & \lambda_{L}=\nn \\
& & \frac{1}{48 \pi^{2}}\left[
6
-66\log\frac{M(3,2,+\tfrac{1}{6})_{\rm VB}}{\mu_{2}}
+3\log\frac{M(3,2,+\tfrac{1}{6})_{\rm GB}}{\mu_{2}}
\right.
\nn\\
& & \qquad
-66\log\frac{M(3,2,-\tfrac{5}{6})_{\rm VB}}{\mu_{2}}
+3\log\frac{M(3,2,-\tfrac{5}{6})_{\rm GB}}{\mu_{2}}
\nn\\
& & \qquad
+3\log\frac{M(3,2,+\tfrac{1}{6})_{\rm CS}^{(2)}}{\mu_{2}}
+3\log\frac{M(3,2,+\tfrac{7}{6})_{\rm CS}^{(2)}}{\mu_{2}}
\nn\\
& & \qquad
+12\log\frac{M(3,3,+\tfrac{1}{3})_{\rm CS}}{\mu_{2}}
+4\log\frac{M(1,3,+1)_{\rm CS}}{\mu_{2}}
\nn\\
& & \qquad
+\tfrac{1}{2}\log\frac{M(1,2,+\tfrac{1}{2})_{\rm RS}^{(1)}}{\mu_{2}}+\tfrac{1}{2}\log\frac{M(1,2,+\tfrac{1}{2})_{\rm RS}^{(2)}}{\mu_{2}}
\nn\\
& &\qquad
+24\log\frac{M(\overline{6},3,-\tfrac{1}{3})_{\rm CS}}{\mu_{2}}
+2\log\frac{M(1,3,0)_{\rm RS}}{\mu_{2}}
\nn\\
& &  \qquad\left.
+8\log\frac{M(8,2,+\tfrac{1}{2})_{\rm CS}^{(2)}}{\mu_{2}}\right] \, ,
\nn
\eea
\bea
& & \lambda_{R}=\nn \\
& & \frac{1}{48 \pi^{2}}\left[
8
-44\log\frac{M(1,1,+1)_{\rm VB}}{\mu_{2}}
+2\log\frac{M(1,1,+1)_{\rm GB}}{\mu_{2}}
\right.
\nn\\
& & \qquad
-66\log\frac{M(3,2,+\tfrac{1}{6})_{\rm VB}}{\mu_{2}}
+3\log\frac{M(3,2,+\tfrac{1}{6})_{\rm GB}}{\mu_{2}}
\nn\\
& & \qquad
-66\log\frac{M(3,2,-\tfrac{5}{6})_{\rm VB}}{\mu_{2}}
+3\log\frac{M(3,2,-\tfrac{5}{6})_{\rm GB}}{\mu_{2}}
\nn\\
& & \qquad
+3\log\frac{M(3,2,+\tfrac{1}{6})_{\rm CS}^{(2)}}{\mu_{2}}
+3\log\frac{M(3,2,+\tfrac{7}{6})_{\rm CS}^{(2)}}{\mu_{2}}
\nn\\
& & \qquad
+\tfrac{1}{2}\log\frac{M(1,2,+\tfrac{1}{2})_{\rm RS}^{(1)}}{\mu_{2}}
+\tfrac{1}{2}\log\frac{M(1,2,+\tfrac{1}{2})_{\rm RS}^{(2)}}{\mu_{2}}
\nn\\
& & \qquad
+12\log\frac{M(6,1,-\tfrac{2}{3})_{\rm CS}}{\mu_{2}}
+8\log\frac{M(8,2,+\tfrac{1}{2})_{\rm CS}^{(2)}}{\mu_{2}}
\nn\\
& & \qquad\left.
+6\log\frac{M(3,1,-\tfrac{4}{3})_{\rm CS}}{\mu_{2}}
+2\log\frac{M(1,1,-2)_{\rm CS}}{\mu_{2}}
\right] \, .
\nn
\eea
%==================================
\paragraph*{The $4_{C}2_{L}1_{R}\to 3_{c}2_{L}1_{R}1_{X}\to 3_{c}2_{L}1_{Y}$ matching at $\mu_{1}$:}
\bea
& & \lambda_{c}=\nn \\
& & \frac{1}{48 \pi^{2}}\left[
1
-22\log\frac{M(3,1,+\tfrac{2}{3})_{\rm VB}}{\mu_{1}}
+\log\frac{M(3,1,+\tfrac{2}{3})_{\rm GB}}{\mu_{1}}
\right.\nn\\
& & \qquad
+2\log\frac{M(3,2,+\tfrac{1}{6})_{\rm CS}^{(3)}}{\mu_{1}}
+2\log\frac{M(3,2,+\tfrac{7}{6})_{\rm CS}}{\mu_{1}}
\nn\\
& &\qquad\left. 
+5\log\frac{M(6,1,+\tfrac{4}{3})_{\rm CS}}{\mu_{1}}
\right]\,,
\nn 
\eea
\bea
& & \lambda_{L}=\frac{1}{16 \pi^{2}}\left[
\log\frac{M(3,2,+\tfrac{1}{6})_{\rm CS}^{(3)}}{\mu_{1}}
+\log\frac{M(3,2,+\tfrac{7}{6})_{\rm CS}}{\mu_{1}}
\right]\,,
\nn 
\eea
\bea
& & \lambda_{RR}=\frac{1}{48 \pi^{2}}\left[2\log\frac{M(1,1,0)_{\rm GB}}{\mu_{1}}+6\log\frac{M(3,1,+\tfrac{2}{3})_{\rm GB}}{\mu_{1}}\right.
\nn \\
& & \qquad
+3\log\frac{M(3,2,+\tfrac{1}{6})_{\rm CS}^{(3)}}{\mu_{1}}
+3\log\frac{M(3,2,+\tfrac{7}{6})_{\rm CS}}{\mu_{1}}
\nn\\
& &\qquad\left. 
+12\log\frac{M(6,1,+\tfrac{4}{3})_{\rm CS}}{\mu_{1}}
\right]\,,
\nn 
\eea
\bea
& & \lambda_{RX}=\lambda_{XR}=-\frac{1}{8\sqrt{6} \pi^{2}}\left[
\log\frac{M(1,1,0)_{\rm GB}}{\mu_{1}}
\right.
\nn \\
& & \qquad
+\log\frac{M(3,1,+\tfrac{2}{3})_{\rm GB}}{\mu_{1}}
-2\log\frac{M(6,1,+\tfrac{4}{3})_{\rm CS}}{\mu_{1}}
\nn\\
& &\qquad
\left. 
+2\log\frac{M(3,2,+\tfrac{1}{6})_{\rm CS}^{(3)}}{\mu_{1}}
+2\log\frac{M(3,2,+\tfrac{7}{6})_{\rm CS}}{\mu_{1}}
\right]\,,
\nn 
\eea
\bea
& & \lambda_{XX}=\nn\\
& & \frac{1}{48 \pi^{2}}\left[
4
-88\log\frac{M(3,1,+\tfrac{2}{3})_{\rm VB}}{\mu_{1}}
+\log\frac{M(3,1,+\tfrac{2}{3})_{\rm GB}}{\mu_{1}}
\right.\nn\\
& & \qquad
+8\log\frac{M(3,2,+\tfrac{1}{6})_{\rm CS}^{(3)}}{\mu_{1}}
+8\log\frac{M(3,2,+\tfrac{7}{6})_{\rm CS}}{\mu_{1}}
\nn\\
& &\qquad\left. 
+2\log\frac{M(6,1,+\tfrac{4}{3})_{\rm CS}}{\mu_{1}}
+3\log\frac{M(1,1,0)_{\rm GB}}{\mu_{1}}
\right]\,.
\nn 
\eea
A few comments are in order. First, a generic analytic identification of the hierarchy of the mass-eigenstates of the mass matrices larger than $2\times 2$ is very difficult due to possible accidental cancellations and, thus, in some cases, the actual mass ordering of the eigenstates labeled by $1,2,3,\ldots$ in the third column of TABLE~\ref{sample82p12} may be different than what is expected from the ordering in the formulas above (where the different eigenstates are labeled by superscripts).\footnote{Notice, for instance, that in the specific example in TABLE~\ref{sample82p12} the mass of $M(3,2,+\tfrac{1}{6})_{\rm CS}^{(2)}$ entering the matching factor $\lambda_{C}$ at  $\mu_{2}$ is smaller than $M(3,2,+\tfrac{1}{6})_{\rm CS}^{(3)}$ entering $\lambda_{c}$ at $\mu_{1}$. } Although this may look contrived, it is readily verified that to a very good accuracy the only effect of such a ``misidentification'' is an overall shift of some of the ``segments'' of the gauge running curves between $\mu_{2}$ and $\mu_{1}$ in figures like Fig.~\ref{samplesolution}, with a negligible effect on the low-energy values of the gauge couplings.\footnote{Indeed, $\log M_{A}/\mu_{1}+\log M_{B}/\mu_{2}=\log M_{A}/\mu_{2}+\log M_{B}/\mu_{1}.$}

Second, we may further simplify the matching formulas by taking advantage of the fact that in the Feynman gauge the masses of the  Goldstones equal those of the corresponding vectors; hence, the contributions from these two sources can be collapsed into a single log term.

Finally, a comment on the artificial ``reality'' of the two SM-Higgs scalars $(1,2,+\tfrac{1}{2})^{(1,2)}_{\rm RS}$ entering the formulas for $\lambda_{L,R}$ at the $\mu_{2}$ matching scale. As a matter of fact, besides the lightest eigenstate playing the role of the SM Higgs doublet, all the heavy complex $(1,2,+\tfrac{1}{2})^{(i)}_{\rm CS}$ eigenstates of the relevant doublet mass matrix should be integrated out at $\mu_{2}$. However, the  spectrum of the doublet-like scalars is unavailable unless all the extra scalars, required for a realistic Yukawa sector (minimally, a complex $10_H$; cf.~also Sect.~\ref{FCNC}), are consistently taken into account. 

Fortunately, such a detailed study is not needed as one can approximate the effect of a light complex doublet by ``averaging" over the eigenvalues of the $2\times 2$ doublet mass matrix in the $45_{H}\oplus 126_{H}$ sector, namely by taking the two ``heavy" doublets as real fields so that the heavy doublet degrees of freedom are counted correctly. As rough as it may sound conceptually, this ``escamotage" has essentially no effect on the numerical results due to the generally very small impact of the scalar doublets on the evolution of the gauge couplings as well as on the associated threshold corrections.  
The interested reader can find a more detailed discussion of these issues in Section~\ref{MSH} and in ref.~\cite{Bertolini:2012im}.
%---------

\section{$SO(10)$ Higgs representations\label{AppC}}

The decomposition of the $10_H$, $45_H$ and $126_H$ representations with respect to all relevant intermediate symmetries is  detailed in TABLEs \ref{tab:10decomp}, \ref{tab:45decomp} and \ref{tab:126decomp}.

\renewcommand{\arraystretch}{1.25}
\begin{table*}[ht]
\begin{tabular}{ccccc|cc}%{llllllllr}
\hline \hline
%$10$ %$SO(10)$
 $4_C\,2_L\,2_R $
& $4_C\,2_L\,1_R $
& $3_c\,2_L\,2_R\,1_{BL} $
& $3_c\,2_L\,1_R\,1_{BL} $
& $3_c\,2_L\,1_Y $
& $5\,1_Z$ %$SU(5)$
& $5'\,1_{Z'}$ %$SU(5)'\otimes U(1)_{Z} $
%& $1_{Y'}$ %$U(1)_{Y'}$
\\
\hline
 $\left({ 6,1,1} \right)$
& $\left({ 6,1},0 \right)$
& $\left({ 3,1,1},-\tfrac{2}{3} \right)$
& $\left({ 3,1},0,-\tfrac{2}{3} \right)$
& $\left({ 3,1},-\tfrac{1}{3} \right)$
& $\left({ 5},-2 \right)$
& $\left({ 5},-2 \right)$
%& $\left. +1 \right.$
\\
\null
& 
& $\left({ \overline{3},1,1},+\tfrac{2}{3} \right)$
& $\left({ \overline{3},1},0,+\tfrac{2}{3} \right)$
& $\left({ \overline{3},1},+\tfrac{1}{3} \right)$
& $\left({ \overline{5}},+2 \right)$
& $\left({ \overline{5}},+2 \right)$
%& $\left. 0 \right.$
\\
$\left({ 1,2,2} \right)$
& $\left({ 1,2},+\frac{1}{2} \right)$
& $\left({ 1,2,2},0 \right)$
& $\left({ 1,2},+\frac{1}{2},0 \right)$
& $\left({ 1,2},+\frac{1}{2} \right)$
& $\left({ 5},-2 \right)$
& $\left({ 5},+2 \right)$
%& $\left. -\frac{5}{6} \right.$
\\
& $\left({ 1,2},-\frac{1}{2} \right)$
& 
& $\left({ 1,2},-\frac{1}{2},0 \right)$
& $\left({ 1,2},-\frac{1}{2} \right)$
& $\left({ \overline{5}},+2 \right)$
& $\left({ \overline{5}},-2 \right)$
\\
\hline \hline
\end{tabular}
\mycaption{Decomposition of the fundamental representation $10$ with respect to the various $SO(10)$ subgroups. 
The definitions and normalization of the abelian charges are given in \sect{45126Higgsmodel}.}
\label{tab:10decomp}
\end{table*}

\renewcommand{\arraystretch}{1.25}
\begin{table*}[ht]
\begin{tabular}{ccccc|cc}%{llllllllr}
\hline \hline
%$10$ %$SO(10)$
 $4_C\,2_L\,2_R $
& $4_C\,2_L\,1_R $
& $3_c\,2_L\,2_R\,1_{BL} $
& $3_c\,2_L\,1_R\,1_{BL} $
& $3_c\,2_L\,1_Y $
& $5\,1_Z$ %$SU(5)$
& $5'\,1_{Z'}$ %$SU(5)'\otimes U(1)_{Z} $
%& $1_{Y'}$ %$U(1)_{Y'}$
\\
\hline
 $\left({ 1,1,3} \right)$
& $\left({ 1,1},+1 \right)$
& $\left({ 1,1,3},0 \right)$
& $\left({ 1,1},+1,0 \right)$
& $\left({ 1,1},+1 \right)$
& $\left({ 10},-4 \right)$
& $\left({ \overline{10}},+4 \right)$
%& $\left. +1 \right.$
\\
\null
& $\left({ 1,1},0 \right)$
&
& $\left({ 1,1},0,0 \right)$
& $\left({ 1,1},0 \right)$
& $\left({ 1},0 \right)$
& $\left({ 1},0 \right)$
%& $\left. 0 \right.$
\\
\null
& $\left({ 1,1},-1 \right)$
&
& $\left({ 1,1},-1,0 \right)$
& $\left({ 1,1},-1 \right)$
& $\left({ \overline{10}},+4 \right)$
& $\left({ 10},-4 \right)$
%& $\left. -1 \right.$
\\
$\left({ 1,3,1} \right)$
& $\left({ 1,3},0 \right)$
& $\left({ 1,3,1},0 \right)$
& $\left({ 1,3},0,0 \right)$
& $\left({ 1,3},0 \right)$
& $\left({ 24},0 \right)$
& $\left({ 24},0 \right)$
%& $\left. 0 \right.$
\\
$\left({ 6,2,2} \right)$
& $\left({ 6,2},+\frac{1}{2} \right)$
& $\left({ 3,2,2},-\frac{2}{3} \right)$
& $\left({ 3,2},+\frac{1}{2},-\frac{2}{3} \right)$
& $\left({ 3,2},+\frac{1}{6} \right)$
& $\left({ 10},-4 \right)$
& $\left({ 24},0 \right)$
%& $\left. -\frac{5}{6} \right.$
\\
\null
& $\left({ 6,2},-\frac{1}{2} \right)$
&
& $\left({ 3,2},-\frac{1}{2},-\frac{2}{3} \right)$
& $\left({ 3,2},-\frac{5}{6} \right)$
& $\left({ 24},0 \right)$
& $\left({ 10},-4 \right)$
%& $\left. +\frac{1}{6} \right.$
\\
\null
&
& $\left({ \overline{3},2,2},+\frac{2}{3} \right)$
& $\left({ \overline{3},2},+\frac{1}{2},+\frac{2}{3} \right)$
& $\left({ \overline{3},2},+\frac{5}{6} \right)$
& $\left({ 24},0 \right)$
& $\left({ \overline{10}},+4 \right)$
%& $\left. -\frac{1}{6} \right.$
\\
\null
&
&
& $\left({ \overline{3},2},-\frac{1}{2},+\frac{2}{3} \right)$
& $\left({ \overline{3},2},-\frac{1}{6} \right)$
& $\left({ \overline{10}},+4 \right)$
& $\left({ 24},0 \right)$
%& $\left. +\frac{5}{6} \right.$
\\
$\left({ 15,1,1} \right)$
& $\left({ 15,1},0 \right)$
& $\left({ 1,1,1},0 \right)$
& $\left({ 1,1},0,0 \right)$
& $\left({ 1,1},0 \right)$
& $\left({ 24},0 \right)$
& $\left({ 24},0 \right)$
%& $\left. 0 \right.$
\\
\null
&
& $\left({ 3,1,1},+\frac{4}{3} \right)$
& $\left({ 3,1},0,+\frac{4}{3} \right)$
& $\left({ 3,1},+\frac{2}{3} \right)$
& $\left({ \overline{10}},+4 \right)$
& $\left({ \overline{10}},+4 \right)$
%& $\left. +\frac{2}{3} \right.$
\\
\null
&
& $\left({ \overline{3},1,1},-\frac{4}{3} \right)$
& $\left({ \overline{3},1},0,-\frac{4}{3} \right)$
& $\left({ \overline{3},1},-\frac{2}{3} \right)$
& $\left({ 10},-4 \right)$
& $\left({ 10},-4 \right)$
%& $\left. -\frac{2}{3} \right.$
\\
\null
&
& $\left({ 8,1,1},0 \right)$
& $\left({ 8,1},0,0 \right)$
& $\left({ 8,1},0 \right)$
& $\left({ 24},0 \right)$
& $\left({ 24},0 \right)$
%& $\left. 0 \right.$
\\
\hline \hline
\end{tabular}
\mycaption{Same as in TABLE~\ref{tab:10decomp} for the $45$ representation.
}
\label{tab:45decomp}
\end{table*}

\renewcommand{\arraystretch}{1.25}
\begin{table*}[ht]
\begin{tabular}{ccccc|cc}%{llllllllr}
\hline \hline
%$10$ %$SO(10)$
 $4_C\,2_L\,2_R $
& $4_C\,2_L\,1_R $
& $3_c\,2_L\,2_R\,1_{BL} $
& $3_c\,2_L\,1_R\,1_{BL} $
& $3_c\,2_L\,1_Y $
& $5\, 1_Z$ %$SU(5)$
& $5'\, 1_{Z'}$
\\
\hline
%$126$
$\left({ 6,1,1} \right)$
& $\left({ 6,1,0} \right)$
& $\left({ \overline{3},1,1},+\frac{2}{3} \right)$
& $\left({ \overline{3},1,0},+\frac{2}{3} \right)$
& $\left({ \overline{3},1},+\frac{1}{3} \right)$
& $\left({ \overline{5}},+2\right)$
& $\left({ \overline{5}},+2\right)$
\\
\null
& 
& $\left({ 3,1,1},-\frac{2}{3} \right)$
& $\left({ 3,1,0},-\frac{2}{3} \right)$
& $\left({ 3,1},-\frac{1}{3} \right)$
& $\left({ 45},-2\right)$
& $\left({ 45},-2\right)$
\\
\null
$\left({ 10,3,1} \right)$
& $\left({ 10,3,0} \right)$ 
& $\left({ 1,3,1},-2 \right)$
& $\left({ 1,3,0},-2 \right)$
& $\left({ 1,3},-1 \right)$
& $\left({ \overline{15}},-6\right)$
& $\left({ \overline{15}},-6\right)$
\\
\null
& 
& $\left({ 3,3,1},-\tfrac{2}{3} \right)$
& $\left({ 3,3,0},-\tfrac{2}{3} \right)$
& $\left({ 3,3},-\tfrac{1}{3} \right)$
& $\left({ 45},-2\right)$
& $\left({ 45},-2\right)$
\\
\null
& 
& $\left({ 6,3,1},+\tfrac{2}{3} \right)$
& $\left({ 6,3,0},+\tfrac{2}{3} \right)$
& $\left({ 6,3},+\tfrac{1}{3} \right)$
& $\left({ \overline{50}},+2\right)$
& $\left({ \overline{50}},+2\right)$
\\
\null
$\left({ \overline{10},1,3} \right)$
& $\left({ \overline{10},1,-1} \right)$ 
& $\left({ 1,1,3},+2 \right)$
& $\left({ 1,1,-1},+2 \right)$
& $\left({ 1,1},0 \right)$
& $\left({ 1},+10\right)$
& $\left({ \overline{50}},+2\right)$
\\
\null
& $\left({ \overline{10},1,0} \right)$ 
& 
& $\left({ 1,1,0},+2 \right)$
& $\left({ 1,1},+1 \right)$
& $\left({ 10},+6\right)$
& $\left({ 10},+6\right)$
\\
\null
& $\left({ \overline{10},1,+1} \right)$ 
& 
& $\left({ 1,1,+1},+2 \right)$
& $\left({ 1,1},+2 \right)$
& $\left({ \overline{50}},+2\right)$
& $\left({ 1},+10\right)$
\\
\null
& 
& $\left({ \overline{3},1,3},+\tfrac{2}{3} \right)$
& $\left({ \overline{3},1,-1},+\tfrac{2}{3} \right)$
& $\left({ \overline{3},1},-\tfrac{2}{3} \right)$
& $\left({ 10},+6\right)$
& $\left({ 45},-2\right)$
\\
\null
& 
& 
& $\left({ \overline{3},1,0},+\tfrac{2}{3} \right)$
& $\left({ \overline{3},1},+\tfrac{1}{3} \right)$
& $\left({ \overline{50}},+2\right)$
& $\left({ \overline{50}},+2\right)$
\\
\null
& 
& 
& $\left({ \overline{3},1,+1},+\tfrac{2}{3} \right)$
& $\left({ \overline{3},1},+\tfrac{4}{3} \right)$
& $\left({ 45},-2\right)$
& $\left({ 10},+6\right)$
\\
\null
& 
& $\left({ \overline{6},1,3},-\tfrac{2}{3} \right)$
& $\left({ \overline{6},1,-1},-\tfrac{2}{3} \right)$
& $\left({ \overline{6},1},-\tfrac{4}{3} \right)$
& $\left({ \overline{50}},+2\right)$
& $\left({ \overline{15}},-6\right)$
\\
\null
& 
& 
& $\left({ \overline{6},1,0},-\tfrac{2}{3} \right)$
& $\left({ \overline{6},1},-\tfrac{1}{3} \right)$
& $\left({ 45},-2\right)$
& $\left({ 45},-2\right)$
\\
\null
& 
& 
& $\left({ \overline{6},1,+1},-\tfrac{2}{3} \right)$
& $\left({ \overline{6},1},+\tfrac{2}{3} \right)$
& $\left({ \overline{15}},-6\right)$
& $\left({ \overline{50}},+2\right)$
\\
\null
$\left({ 15,2,2} \right)$
& $\left({ 15,2,-\tfrac{1}{2}} \right)$ 
& $\left({ 1,2,2},0 \right)$
& $\left({ 1,2,-\tfrac{1}{2}},0 \right)$
& $\left({ 1,2},-\tfrac{1}{2} \right)$
& $\left({ \overline{5}},+2\right)$
& $\left({ 45},-2\right)$
\\
\null
& $\left({ 15,2,+\tfrac{1}{2}} \right)$ 
&
& $\left({ 1,2,+\tfrac{1}{2}},0 \right)$
& $\left({ 1,2},+\tfrac{1}{2} \right)$
& $\left({ 45},-2\right)$
& $\left({ \overline{5}},+2\right)$
\\
\null
&
& $\left({ \overline{3},2,2},-\tfrac{4}{3} \right)$
& $\left({ \overline{3},2,-\tfrac{1}{2}},-\tfrac{4}{3} \right)$
& $\left({ \overline{3},2},-\tfrac{7}{6} \right)$
& $\left({ 45},-2\right)$
& $\left({ \overline{15}},-6\right)$
\\
\null
&
& 
& $\left({ \overline{3},2,+\tfrac{1}{2}},-\tfrac{4}{3} \right)$
& $\left({ \overline{3},2},-\tfrac{1}{6} \right)$
& $\left({ \overline{15}},-6\right)$
& $\left({ 45},-2\right)$
\\
\null
&
& $\left({ 3,2,2},+\tfrac{4}{3} \right)$
& $\left({ 3,2,+\tfrac{1}{2}},+\tfrac{4}{3} \right)$
& $\left({ 3,2},+\tfrac{7}{6} \right)$
& $\left({ \overline{50}},+2\right)$
& $\left({ 10},+6\right)$
\\
\null
&
& 
& $\left({ 3,2,-\tfrac{1}{2}},+\tfrac{4}{3} \right)$
& $\left({ 3,2},+\tfrac{1}{6} \right)$
& $\left({ 10},+6\right)$
& $\left({ \overline{50}},+2\right)$
\\
\null
& 
& $\left({ 8,2,2},0 \right)$
& $\left({ 8,2,-\tfrac{1}{2}},0 \right)$
& $\left({ 8,2},-\tfrac{1}{2} \right)$
& $\left({ \overline{50}},+2\right)$
& $\left({ 45},-2\right)$
\\
\null
& 
&
& $\left({ 8,2,+\tfrac{1}{2}},0 \right)$
& $\left({ 8,2},+\tfrac{1}{2} \right)$
& $\left({ 45},-2\right)$
& $\left({ \overline{50}},+2\right)$
\\
\hline \hline
\end{tabular}
\mycaption{Same as in TABLE~\ref{tab:10decomp} for the $126$ representation.}
\label{tab:126decomp}
\end{table*}

\end{document}